\documentclass[prx, preprint, superscriptaddress]{revtex4-2}
\usepackage{amsmath,hyperref,cleveref,amssymb,amsfonts,graphicx, multirow, color, bm,textcomp,mathtools}
\usepackage{paralist}
\usepackage[all,cmtip]{xy}
\usepackage{mathrsfs}
\usepackage{soul, subfigure}
\usepackage{threeparttable,dsfont,bbm}
\usepackage{color}
\usepackage{comment}
\usepackage{dcolumn}
\usepackage{array,tabularx}
\usepackage{booktabs}

\usepackage{adjustbox}

\def\kv{{\bf k}}
\def\av{{\bf a}}

\begin{document}

\title{Altermagnetism, Kagome Flat Band, and Weyl Fermion States in Magnetically Intercalated Transition Metal Dichalcogenides}

\author{Avinash Sah}
\affiliation {Department of Physics and Astronomy, University of Missouri, Columbia, Missouri 65211, USA}

\author{Ting-Yong Lim }
\affiliation {Department of Physics, National Cheng Kung University, Tainan 701, Taiwan}

\author{Clayton Conner}
\affiliation {Department of Physics and Astronomy, University of Missouri, Columbia, Missouri 65211, USA}

\author{Amarnath Chakraborty}
\affiliation {Department of Physics and Astronomy, University of Missouri, Columbia, Missouri 65211, USA}

\author{Giovanni Vignale}
\affiliation {Department of Physics and Astronomy, University of Missouri, Columbia, Missouri 65211, USA}

\author{Tay-Rong Chang}
\affiliation {Department of Physics, National Cheng Kung University, Tainan 701, Taiwan}
\affiliation {Center for Quantum Frontiers of Research and Technology (QFort), Tainan, 70101, Taiwan}
\affiliation {Physics Division, National Center for Theoretical Sciences, Taipei, 10617, Taiwan}

\author{Pavlo Sukhachov}\email{Corresponding Author: Pavlo Sukhachov, pavlo.sukhachov@missouri.edu}
\affiliation {Department of Physics and Astronomy, University of Missouri, Columbia, Missouri 65211, USA}
\affiliation {MU Materials Science \& Engineering Institute, University of Missouri, Columbia, Missouri 65211, USA}

\author{Guang~Bian}\email{Corresponding Author: Guang Bian, biang@missouri.edu}
\affiliation {Department of Physics and Astronomy, University of Missouri, Columbia, Missouri 65211, USA}
\affiliation {MU Materials Science \& Engineering Institute, University of Missouri, Columbia, Missouri 65211, USA}

\newpage
\begin{abstract}
{\bf Altermagnetic (AM) compounds have recently emerged as a promising platform for realizing unconventional quantum phases, enabled by their unique spin-split band structure at zero net magnetization. Here, we present a first-principles investigation of magnetically intercalated transition metal dichalcogenides (TMDs) of the form XY$_4$Z$_8$ (X $=$ Mn, Fe, Co, Ni, Cr, or V; Y $=$ Nb or Ta; and Z $=$ Se or S), identifying a subset of new versatile AM candidates. Our results establish a systematic correlation between interatomic geometry, quantified by the ratio of interlayer to intralayer spacing, and the magnetic ground states. Systems with A-type antiferromagnetic order exhibit momentum-dependent spin splitting consistent with AM behavior. The combination of the AM spin-splitting and the spin-orbit coupling leads to the emergence of Weyl nodes together with the corresponding topological Fermi arc surface states. Moreover, we identify flat bands near the Fermi level that originate from the intercalant-induced formation of an effective kagome-like sublattice in the TMD layer. These results collectively establish magnetically intercalated TMDs as a promising platform for engineering altermagnetism, flat bands, and Weyl fermions within a single material family, facilitating the development of topological and spintronic applications.}
\end{abstract}


\maketitle

\newpage
\section*{Introduction}

Antiferromagnetic materials (AFMs) represent one of the fundamental classes of magnetic materials with antiparallel spin configurations. Recently, the field of magnetism has enjoyed a renewal of interest related to spin-split AFMs. The original ideas explaining momentum-dependent spin-splitting, i.e., the mixing of coordinate and spin degrees of freedom in the absence of spin-orbital coupling (SOC)~\cite{Pekar-Rashba-CombinedResonanceCrystals-1965}, were proposed several decades ago. However, symmetry classification in terms of the spin groups~\cite{Smejkal-Sinova:2020, Smejkal-Jungwirth:2022} and \textit{ab initio} characterization~\cite{Noda-Nakamura-MomentumdependentBandSpin-2016, Smejkal-Sinova:2020, Hayami-Kusunose-MomentumDependentSpinSplitting-2019, Yuan-Zunger:2020, Yuan-Zunger-PredictionLowZCollinear-2021, Ma-Liu-MultifunctionalAntiferromagneticMaterials-2021, Smejkal-Jungwirth:2022, Smejkal-Jungwirth-ConventionalFerromagnetismAntiferromagnetism-2022} were provided only relatively recently. Unlike the cases of conventional AFMs, the symmetry operations of spin-split AFMs (without translation or inversion operations) lead to a lifted spin degeneracy away from the spin-degenerate lines or planes in the AFM ground state, giving rise to the so-called altermagnetic (AM) phase. The spin-splitting of the band structure of altermagnets has even parity, i.e., it has $d$-, $g$-, $i$-, etc. structure, drawing a link to the symmetry of the order parameter in superconductors~\cite{Jungwirth-Smejkal-AltermagnetismUnconventionalSpinordered-2025}. Up to date, a multitude of AM material candidates, such as MnF$_2$, ultra-thin films of RuO$_2$, Mn$_5$Si$_3$, MnTe, V$_2$Te$_2$O, KV$_2$Se$_2$O, Rb$_{1-\delta}$V$_2$Te$_2$O, CrO, and CrSb, have been suggested. Please see Refs.~\cite{Smejkal-Jungwirth:2022, Bai-Yao-AltermagnetismExploringNew-2024} for a more comprehensive list of altermagnetic candidates, and Refs.~\cite{Song-Pan-AltermagnetsNewClass-2025, Jungwirth-Smejkal-AltermagnetismUnconventionalSpinordered-2025, Jungwirth-Smejkal-SymmetryMicroscopySpectroscopy-2025} for recent reviews. Signatures of spin-split electron bands were experimentally observed in Refs.~\cite{Krempasky-Jungwirth:2024, Osumi-Sato-ObservationGiantBand-2024, Reimers-Jourdan:2023, Zeng-Liu-ObservationSpinSplitting-2024, Ding-Shen-LargeBandsplittingWave-2024, Jeong-Jalan-AltermagneticPolarMetallic-2024, Betancourt2024, Weber-Schneider-AllOpticalExcitation-2024, Jiang-Qian-MetallicRoomtemperatureDwave-2025, Zhang-Chen-CrystalsymmetrypairedSpinValley-2025} by using spectroscopic and transport measurements.

Recently, a promising platform for altermagnetism has emerged in transition-metal dichalcogenides (TMDs) intercalated with magnetic atoms. TMDs are layered van der Waals materials that have long attracted attention due to their structural flexibility, orbital richness, and the ease with which their dimensionality, carrier density, and symmetry can be tuned by intercalation, gating, or strain~\cite{Wilson-DiSalvo-DW-TMDs-1975, Manzeli-TMDsReview-2017, Musfeldt2025CrTaS2}. This versatility has already enabled the discovery of charge density waves~\cite{Wilson-DiSalvo-DW-TMDs-1975, Rossnagel-CDWTMDs-2011}, unconventional superconductivity~\cite{Saito-Iwasa-IsingSC-2016, Xi-Cao-IsingSCNbSe2-2016}, and correlated insulating states in pristine and doped TMDs~\cite{Kushwaha2025CrTeMonolayers, Wang-Dean-CorrelatedElectronicPhases-2020, Chen-Crommie-StrongCorrelationsOrbital-2020}, establishing them as a playground for emergent quantum phases. When magnetic atoms are intercalated into van der Waals gaps, they provide an additional knob to couple spin, orbital, and lattice degrees of freedom, thereby reshaping both the symmetry and electronic structure. Within this context, magnetically intercalated TMD compounds constitute a particularly compelling and largely unexplored material class. Interestingly, the crystal structure of the XY$_4$Z$_8$ (or X$_{0.25}$YZ$_2$, where X is the magnetic intercalant and YZ$_2$ is a TMD compound) composition stabilizes a layered AM phase, supports a layered AM phase, driven by the interplay between lattice symmetry and antiparallel spin ordering. For example, by combining experimental methods and \textit{ab initio} calculations, CoNb$_4$Se$_8$ has recently been identified as one of the first representatives for the realization of this broader physics, exhibiting $g$-wave altermagnetism~\cite{regmi1, dale2024nonrelativisticspinsplittingfermi, sakhya2025electronicstructurelayeredaltermagnetic, opticalswitch2025, candelora2025discoveryintertwinedspincharge}. Beyond altermagnetism, these intercalated compounds offer a natural setting where gapped altermagnetic nodal lines evolve into SOC-stabilized Weyl fermions, and where kagome-derived flat bands manifest through orbital localization and destructive interference. Yet, the other members of the XY$_4$Z$_8$ family, which were long regarded as conventional AFMs, remain largely unexplored~\cite{PhysRevB.107.184429}. In particular, the potential realization of altermagnetism in these AFM compounds, together with nontrivial band topologies, has not been systematically investigated, leaving open a unique opportunity to unify altermagnetism and band topology within a single material platform.

In this work, we perform a comprehensive \textit{ab initio} characterization of all XY$_4$Z$_8$ compounds, where X = $\text{Mn, Fe, Co, Ni, Cr, V}$, Y = $\text{Nb, Ta}$, and Z = $\text{Se, S}$. Our analysis reveals a direct correlation between interatomic geometry quantified by the ratio of interlayer to intralayer spacing and magnetic ground state selection. In addition to CoNb$_4$Se$_8$, materials such as FeNb$_4$Se$_8$, CoTa$_4$Se$_8$, and FeNb$_4$S$_8$ are identified as altermagnets. Remarkably, it is shown that magnetic intercalation promotes weakly dispersive flat-band features near the Fermi energy, resulting from an effective kagome-like lattice. Furthermore, the combination of altermagnetic spin splitting and SOC allows these materials to host Weyl fermions and topological Fermi arc surface states. In other words, the magnetically intercalated TMD compounds XY$_4$Z$_8$ simultaneously support layered antiferromagnetic order and an effective kagome-like lattice geometry, creating a natural setting for momentum-dependent spin splitting, orbital-selective band flattening, and nontrivial band topology to emerge through a unified interplay of lattice geometry, orbital hybridization, and magnetic order. Therefore, this class of materials provides a unique platform that hosts a wide spectrum of quantum phases, including altermagnetism, flat-band states, and topological Weyl fermions.

\section*{Results \& Discussion}

\subsection*{Relation between the lattice geometry and magnetic orderings}
First, we present a comprehensive DFT analysis to examine the emergence of altermagnetism and orbital-selective band features in XY$_4$Z$_8$ systems driven by the interplay of atomic composition, geometric structure, and magnetic ordering. The crystal structures of all relaxed intercalated compounds retain the original hexagonal symmetry (space group P$6_3/mmc$, No.~194) and are visualized in Fig.~1\textbf{a,b}. The side view in Fig.~1\textbf{a} highlights the stacking of chalcogen–metal–chalcogen trilayers, where two intercalant atoms (X) are located in van der Waals (vdW) gaps between adjacent 2H-TMD layers. This intercalated geometry is a well-established characteristic of layered TMDs, stabilized by weak interlayer van der Waals forces~\cite{Chhowalla2013, Conner2025CrTe2}. The key geometric parameters in Fig.~1\textbf{a} are the interlayer spacing ($u$) and the intralayer spacing ($v$), which are defined as the vertical and in-plane distances between the two intercalated atoms per unit cell. The top view in Fig.~1\textbf{b} further provides a planar view of the intercalant positions. Fig.~1\textbf{c} schematically shows the AFM and FM spin arrangements, while Fig.~1\textbf{d} presents the 3D Brillouin zone and symmetry paths used in our band-structure calculations.

Intercalant atoms within the vdW gaps give rise to two distinct classes of exchange pathways that compete to determine the magnetic ground state. Intralayer exchange within the van der Waals gap is dominated by carrier-mediated RKKY-type interactions. Intercalants are laterally separated with no short shared-ligand bridges; therefore, any direct superexchange-like process within the plane would require long, multi-step virtual hopping through the Nb–Se network and is therefore strongly suppressed.  Instead, the dominant mechanism is the polarization of itinerant Nb-derived conduction electrons by localized intercalant moments, which mediates an indirect RKKY-like coupling. As the in-plane susceptibility of the layered NbSe$_2$ host is relatively large and the relevant intercalant separations fall within the short-distance regime of the RKKY interaction, this intralayer coupling generally favors ferromagnetic alignment within each intercalant plane. Consequently, the magnetic ground state is primarily determined by how these ferromagnetically aligned planes stack along the out-of-plane direction. 
Interlayer exchange couples intercalants in adjacent van der Waals gaps separated by a distance $u$. Each intercalant (X) is locally coordinated by chalcogen ligands both above and below, allowing efficient hybridization between X $d$ orbitals and Se $p$ states, which in turn couple strongly to Nb $d$ states in adjacent layers. This vertical connectivity supports higher-order superexchange-like virtual hopping through $X$–Se–Nb–Se–$X$ paths. In addition, carrier-mediated RKKY exchange through itinerant Nb electrons is also present. The resulting interlayer interaction therefore reflects a competition between a superexchange-like term that becomes more effective as the vertical separation is reduced and an electronic-structure-dependent interlayer RKKY contribution whose magnitude and sign are governed by the conduction-electron spin susceptibility. Importantly, the RKKY-type interaction remains operative in both channels: it robustly stabilizes ferromagnetic alignment within each intercalant plane, while its interlayer component can either reinforce or oppose the superexchange-like contribution depending on the Fermi wave vector ($k_{\mathrm F}$) and the distance between the magnetic ions. 

Within this microscopic picture, the $u/v$ ratio provides a useful structural metric for gauging the relative leverage of interlayer versus intralayer coupling. Our DFT results reveal a systematic correlation between the $u/v$ ratio and the preferred magnetic ground state, as summarized in Table~\ref{Table1}. When $u/v < 0.91$, the $X$–Nb–$X$ bond angles are smaller, pulling the intercalants closer together along the out-of-plane direction. This reduced vertical separation enhances orbital overlap mediated by the central host atom (Nb) and the intervening chalcogen ligands. In this regime, the increased strength of interlayer coupling is sufficient to overcome the intralayer tendency toward ferromagnetic alignment and favors antiparallel stacking of ferromagnetically aligned planes. As a result, compounds in this regime tend to stabilize in an AFM ground state. Such bond-angle-controlled superexchange mechanisms are well described by the Goodenough–Kanamori–Anderson (GKA) rules and have been discussed in various systems in Refs.~\cite{Goodenough1955, Kanamori1959, Ovanesyan1975, Sadhukhan2022}. Consistent with this trend, we identify four compounds, FeNb$_4$S$_8$, FeNb$_4$Se$_8$, CoNb$_4$Se$_8$, and CoTa$_4$Se$_8$, that stabilize in A-type AFM ground states. The magnetic ordering in these systems is in close agreement with earlier experimental and theoretical works in Refs.~\cite{PhysRevB.110.144420, doi:10.1021/acs.inorgchem.3c02652, Kouarta2019, PhysRevB.107.184429}. In contrast, an increase in the $X$–Nb–$X$ bond angle leads to $u/v \ge 0.91$, resulting in a more laterally extended geometry and an increased interlayer spacing between intercalants. This structural relaxation reduces orbital overlap and the overall effectiveness of vertical exchange pathways, thereby weakening interlayer coupling. With reduced vertical coupling, magnetic exchange becomes increasingly dominated by in-plane pathways, where long-range RKKY-like interaction mediated by itinerant carriers is more effective. Such competing magnetic interactions, particularly the role of long-range RKKY-type coupling in intercalated TMDs, play a significant role in determining magnetic ordering~\cite{Parkin1980, doi:10.1021/jacs.1c12975}. This enhanced itinerant exchange stabilizes a ferromagnetic ground state in compounds with larger $u/v$ ratios.

The relative strength of the two interlayer contributions evolves across the series with $d$-shell filling. Superexchange terms rely on low-energy hopping pathways, which become progressively reduced as the $d$ shell fills due to Pauli blocking and higher excitation energies. In contrast, the carrier-mediated (RKKY-like) component is governed by the coupling between local moments and itinerant Nb electrons. As a result, changing the intercalant can substantially renormalize, and in borderline cases even reverse, the effective interlayer coupling. While the absolute values of the in-plane and out-of-plane lattice parameters may vary slightly depending on the choice of exchange–correlation functional or structural relaxation protocol, the relative energetic stability between competing magnetic configurations remains robust over a broad range of on-site Coulomb interaction values across the family of altermagnetic candidates (see Supplementary Table~T2). This demonstrates that the magnetic ground state is insensitive to moderate variations in correlation strength.

The only notable exception to the systematic trend in the $u/v$ ratio is NiNb$_4$Se$_8$, which, despite having a ratio below 0.91, stabilizes in a ferromagnetic phase. This exception arises from the nearly filled $d$-orbital configuration of Ni, which fundamentally alters the exchange mechanisms. The near-full occupation of the $d$ orbitals suppresses interlayer superexchange pathways even though the geometric interlayer distance is relatively small (see Supplementary Table~T4 for quantitative energy differences under strain). At the same time, increased $d$-electron filling shifts the $d$-derived conduction bands near the Fermi level, modifying $k_{\mathrm F}$ and enhancing the sensitivity of the itinerant RKKY-like interaction. The borderline nature of this compound is evidenced by the remarkably small energy difference between FM and A-AFM configurations compared to those of the Fe and Co compounds (see Supplementary Tables~T2 and T4). This near degeneracy reflects a delicate balance between weakened interlayer superexchange and renormalized itinerant exchange, allowing ferromagnetic ordering to emerge. Recent theoretical work likewise indicates that this compound resides near the boundary between antiferromagnetic and ferromagnetic order, making its ground state highly sensitive to subtle variations in electronic correlations and spin–orbit coupling~\cite{Gong2025}. Magnetic ordering trends across the series of intercalated TMDs are summarized in Table~\ref{Table1}. Together, these trends highlight the importance of orbital symmetry, bonding character, and lattice anisotropy in determining magnetic order.

\subsection*{Spin splitting in the band structure of altermagnetic \texorpdfstring{$\mathrm{XY}_4\mathrm{Z}_8$}{XY4Z8} compounds}

As shown in Table~\ref{Table1}, the magnetic ground states of intercalated TMDs are influenced by the balance between interlayer and intralayer interactions. These structural and magnetic variations play a critical role in shaping the electronic structure, particularly in determining the presence or absence of spin degeneracy across various planes. In this context, compounds exhibiting A-type AFM configurations, where spins align ferromagnetically within individual layers but antiferromagnetically between adjacent layers, emerge as altermagnetic candidates. This layered magnetic arrangement leads to a compensated magnetic structure with zero net magnetization per unit cell, yet the combination of magnetic ordering and crystal symmetry breaking permits momentum-dependent spin polarization. A characteristic feature of these structures is the relative orientation of the two magnetic sublayers. Although crystallographically equivalent, these layers are rotated by $180^\circ$ due to the P$6_3/mmc$ stacking symmetry in Figs.~1\textbf{a,b}. This rotation, coupled with antiparallel spin alignment as seen in the left panel of Fig.~1\textbf{c}, breaks global spin-rotation and time-reversal symmetry while preserving their combined operation. In a conventional collinear AFM, the time reversal and translation symmetry maps one spin sublattice onto the other and thereby enforces spin degeneracy at all $\boldsymbol{k}$ points in the Brillouin zone. By contrast, in altermagnets this degeneracy protection is absent in off-nodal planes, and the relevant spin group permits momentum-dependent spin splitting even in the absence of net magnetization - a hallmark of altermagnets~\cite{Smejkal-Sinova:2020, Smejkal-Jungwirth:2022}.

Within the spin space-group (SSG) framework, applicable in the absence of SOC, symmetry operations take the general form $\left[R_s \mid \mid R_r\right]$, where $R_r$ is a real-space symmetry and $R_s \in \mathrm{SO}(3)$ represents an independent spin rotation. The numeration and the representation theory of SSG are provided in Refs.~\cite{Litvin-SpinPointGroups-1977, Chen-Liu-EnumerationRepresentationTheory-2024, Cheong2025AltermagnetismClassification}, and these groups played a crucial role in the classification of altermagnets~\cite{Smejkal-Jungwirth-ConventionalFerromagnetismAntiferromagnetism-2022}.
For the present A-type AFM structure, the magnetic order is described by the altermagnetic spin space group $P^{-1}6_3/^{-1}m^1m^{-1}c^{\infty m}1$ in the Belov-Neronova-Smirnova SSG notations~\cite{FINDSPINGROUP}. The SSG describes real-space symmetries including the sixfold screw axis $6_3$, mirror planes $m$, and spin rotation axis $c$ that, when combined with appropriate spin rotations, leave the A-type AFM magnetic configuration invariant. In particular, the opposite-spin sublattices are related by real-space symmetries that exchange the two magnetic layers. Particularly, the mirror symmetry $M_z$ and the sixfold screw symmetry $C_{6z}t$ both interchange the magnetic sublayers. Within the SSG framework, these operations are combined with a $\pi$ spin rotation to form the symmetry operations $[C_2 \parallel M_z]$ and $[C_2 \parallel C_{6z}t]$, which connect the opposite-spin sublattices and preserve the magnetic configuration. Such SSG symmetries allow momentum-dependent spin splitting without net magnetization, with degeneracy enforced on symmetry-invariant nodal planes. When SOC is included, spin rotations no longer decouple from the lattice rotations, and the SSG description reduces to the corresponding magnetic space-group (MSG) framework. In this case, A-type AFM stacking lowers the symmetry of the parent space group $P6_3/mmc$ to the type-III MSG $P6_3'/m'm'c$, in which time-reversal symmetry is broken but specific antiunitary combinations of crystal symmetries with $\mathcal{T}$ remain preserved; the latter MSG agrees with that suggested in Ref.~\cite{regmi1}.

In our system, the protection of spin degeneracy in the nodal plane at $k_z = 0$ arises from the mirror symmetry combined with the spin rotation $[C_2 \parallel M_z]$ that leaves the $k_z = 0$ plane invariant. This results in a Kramers-like twofold degeneracy throughout the entire $k_z = 0$ plane, enforcing spin-degenerate nodal lines along high-symmetry paths such as $\Gamma$--M--K--$\Gamma$, as shown in Fig.~1(d). This symmetry protection is evident in the spin-resolved band structures of representative altermagnetic compounds, including CoNb$_4$Se$_8$, FeNb$_4$Se$_8$, CoTa$_4$Se$_8$, and FeNb$_4$S$_8$ [Figs.~2(a--d), left], where spin-up and spin-down bands remain degenerate along $\Gamma$--M--K--$\Gamma$. Similar symmetry-protected nodal-plane degeneracies have been reported in MnTe and CrSb altermagnetic compounds. In contrast, in the off-nodal plane at $k_z = \pi/2c$, the mirror symmetry $M_z$ no longer leaves momenta within the same plane invariant, allowing for momentum-dependent spin splitting. The latter, however, is still restricted by the combined $C_{6z}$-symmetric mirror planes and spin rotation leading to the $g$-wave symmetric altermagnetic nodes. This behavior is illustrated in the right panels of Figs.~2(a--d), where pronounced spin splitting appears along the low-symmetry $\Gamma'$--M' path in the off-nodal plane at $k_z = \pi/2c$ (see Supplementary Fig.~S1 for calculations without SOC).

To further highlight this behavior, we computed the Fermi surfaces with and without SOC for representative altermagnets. The results for CoNb$_4$Se$_8$ and FeNb$_4$Se$_8$ are shown in Figs.~3\textbf{a-d}. For completeness, the corresponding Fermi surfaces of CoTa$_4$Se$_8$ and FeNb$_4$S$_8$ are provided in Supplementary Fig.~\textbf{S2}, which confirms that the same qualitative features persist across different members of the altermagnet family. The left panel, corresponding to the $k_z = 0$ plane, shows complete spin degeneracy across the Brillouin zone, in line with the symmetry-protected nodal lines discussed above. By contrast, the right panel at $k_z = \pi/2c$ clearly exhibits pronounced spin splitting and strong spin polarization along low-symmetry paths such as $\Gamma^{\prime}$-M$^{\prime}$, consistent with the band dispersions in Fig.~2. This direct comparison shows that, in A-type AFM systems, spin splitting originates from symmetry breaking even without SOC, as was proposed in altermagnets~\cite{PhysRevX.12.011028}. With SOC, although there is band splitting, the symmetry constraints enforce degeneracy at high-symmetry nodal planes. At off-nodal planes where the symmetry protections are absent, the SOC enhances the spin splitting. Among the investigated altermagnetic compounds, CoTa$_4$Se$_8$ exhibits the largest spin-split band magnitude, reaching approximately 110 meV at the Fermi level and increasing to about 154 meV at $-0.25$ eV. The remaining altermagnetic compounds display smaller but qualitatively similar momentum-dependent spin splitting. A complete summary of the spin-splitting magnitudes across all compounds is provided in Supplementary Table~T1. These theoretical results agree with both ARPES and $\mu$SR measurements in CoNb$_4$Se$_8$, which support its altermagnetic nature in Refs.~\cite{dale2024nonrelativisticspinsplittingfermi,sakhya2025electronicstructurelayeredaltermagnetic, graham2025localprobeevidencesupporting}.

In contrast to altermagnetic compounds, the FM-ordered systems MnNb$_4$Se$_8$, NiNb$_4$Se$_8$, CrNb$_4$Se$_8$, and VNb$_4$Se$_8$ exhibit spin splitting that originates from their net magnetic moments. This Zeeman-like effect lifts spin degeneracy uniformly across the Brillouin zone, producing a nearly momentum-independent band separation characteristic of conventional ferromagnetism. The spin-resolved electronic band structures of these FM compounds are provided in Supplementary Fig.~\textbf{S3}, which shows the dispersion along the $\Gamma$–M–K–$\Gamma$ path. In each case, a global spin polarization is evident, consistent with earlier reports on related intercalated ferromagnets such as NiTa$_4$Se$_8$ and MnTa$_4$S$_8$~\cite{PhysRevB.106.224429, PhysRevResearch.4.013048}. The spin splitting observed in altermagnetic TMDs contrasts sharply with their FM counterparts. This distinction highlights the unique role of crystal symmetry and magnetic stacking in enabling spin splitting in altermagnets. As a result, intercalated TMDs with altermagnetic order present a compelling platform for realizing altermagnetic behavior, where symmetry-enabled spin polarization arises in the absence of net magnetization. Moreover, recent studies suggest that such altermagnetic TMDs may host intertwined charge and spin density waves, enable ultrafast optical switching, and even support higher-order topological superconductivity, see Refs.~\cite{candelora2025discoveryintertwinedspincharge, opticalswitch2025, PhysRevB.109.224502}.

\subsection*{Weyl fermion states and Fermi arc surface states in AM compounds}
\setlength{\parskip}{6pt}
\setlength{\parindent}{14pt}
The altermagnetic spin texture in A-type AFM, while intrinsically nonrelativistic, provides the platform for realizing a Weyl semimetal in an AFM ground state. Previously, Weyl semimetals have been extensively studied only in the nonmagnetic and ferromagnetic compounds, as the conventional AFM compounds have no spin splitting in the band structure. Altermagnetic compounds with spin-split bands open new avenues for investigating topological responses of Weyl fermions—such as the chiral anomaly—within a more magnetic-field-tolerant environment.

Our calculations reveal CoNb$_4$Se$_8$ as a Weyl semimetal whose bulk Weyl points (WPs) emerge from SOC-gapped altermagnetic nodal lines. In AM compounds, the alternating spin polarization due to altermagnetism enforces opposite-spin band crossings that extend along momentum-dependent nodal lines in the absence of SOC. These nodal manifolds arise because the spin-up and spin-down channels are decoupled, yet symmetry-related, allowing their dispersions to remain degenerate without hybridization. When SOC is included, it couples both spin channels and gaps most of the degeneracies along the nodal manifold. However, crystalline antiunitary symmetry operations that combine spatial symmetry with time reversal prevent a complete removal of degeneracy, leaving behind residual degeneracies at isolated momenta~\cite{Murakami2007NJP, Soluyanov2015Nature, Wan:2011udc, Armitage2018RMP, Liu2022Co3Sn2S2, GMSS:book}. As a result, the continuous nodal lines collapse into discrete band crossing points. These crossing points represent bulk WPs that emerge as remnants of altermagnetic nodal lines. The distribution of these WPs across the Brillouin zone is mapped in Fig.~4\textbf{a}, revealing 12 pairs of WPs ($\chi = +1$ in dashed red and $\chi = -1$ in dashed blue) within $\pm 0.2$ eV of the Fermi energy, displaced along low-symmetry directions. This anisotropic distribution reflects inversion-symmetry breaking in the A-type AFM stacking and demonstrates the momentum-selective character of altermagnetic spin splitting. Similar SOC-driven conversion of nodal lines into WPs has been recently reported in CrSb~\cite{Li2025CrSb, Lu2025CrSb}.

The microscopic nature of these SOC-induced Weyl points is illustrated in Fig.~4\textbf{b}. The top-left panel provides a zoomed view of the surface spectral function along the $K$-path, where SOC lifts the nodal degeneracies and leaves residual degeneracies as isolated Weyl points. The top right panel displays the corresponding band dispersion along the same path, where two WPs are clearly resolved at an energy of $E = 0.17$~eV. These features directly reflect the SOC-induced splitting of the altermagnetic nodal crossings and mark the emergence of a topological semimetallic phase. The lower three panels show the dispersions of the Weyl cones along orthogonal momentum cuts ($k_x$, $k_y$, and $k_z$) through the WP locations. Each cut reveals the linear energy-momentum relationship characteristic of Weyl fermions, confirming their anisotropic yet gapless cone-like structure.

The surface correspondence of these Weyl nodes is shown in Fig.~4\textbf{c}. In the main panel, momentum-resolved spectral weight highlights non-trivial surface dispersions characteristic of Weyl semimetals, manifesting as surface Fermi arcs (SFAs) that terminate precisely at the projected bulk Weyl nodes with opposite topological charge~\cite{Wan:2011udc, Haldane:2014, Xu2015TaAs}, thereby enforcing the bulk-boundary correspondence associated with nonvanishing Chern flux. Importantly, in the surface Brillouin zone, the projection of two bulk WPs with the same chirality onto a common point gives rise to double Fermi arcs. These arcs arise from surface-projection overlap rather than from genuine bulk double Weyl points. The enlarged inset in Fig.~4\textbf{c} highlights the connectivity of one such double Fermi arc segment (see Supplementary Fig.~\textbf{S5} for robustness check of Fermi arcs against surface variations). Finally, the topological charge of bulk WPs is quantified in Fig.~4\textbf{d}, where the evolution of spin-resolved hybrid Wannier charge centers around the bulk WPs yields chiral charges $\chi=+1$ and $\chi=-1$, confirming the presence of a non-trivial Berry curvature flux. Overall, the absence of trivial backfolded surface contours, together with the open arc topology, provides a definitive spectroscopic fingerprint of Weyl quasiparticles in this system. Experimentally, these Weyl points can be accessible via moderate electron doping through an access intercalant. In CoNb$_4$Se$_8$, excess Co acts as an electron donor through Co-Nb hybridization, shifting the chemical potential upward, as established in Co-intercalated NbSe$_2$ systems. Notably, the $2 \times 2$ intercalation structure remains stable in A-type AFM ground state even at elevated Co concentrations, indicating robustness against moderate enrichment of Co~\cite{Mandujano2025_JACS147_44926}. This establishes CoNb$_4$Se$_8$ as a Weyl semimetal whose topological properties are governed by the cooperative action of altermagnetic spin splitting and SOC, giving rise to SOC-stabilized  WPs. 

\subsection*{Effective kagome lattice in \texorpdfstring{$\mathrm{XY}_4\mathrm{Z}_8$}{XY4Z8} compounds}

In addition to altermagnetic spin textures and Weyl fermions, a remarkable feature consistently observed across several intercalated compounds is the emergence of coexisting flat bands and Dirac-like crossings centered at the high-symmetry $K$ point, as shown in Fig.~5\textbf{a}. These features reflect a delicate interplay between orbital character, intercalation-induced structural modification, and the symmetry properties of the underlying TMD host lattice. To uncover the microscopic origin of these features, we performed atomic and orbital-resolved electronic structure calculations for CoNb$_4$Se$_8$. Figure~5\textbf{a} presents the atom-projected band structure of CoNb$_4$Se$_8$, where contributions from Co (cyan), Nb (magenta), and Se (brown) are highlighted. Figures~5\textbf{b} and 5\textbf{c} provide an orbital decomposition of the Nb $4d$ orbitals, where Fig.~5\textbf{b} corresponds to the central Nb atom (Site 1) located directly beneath the Co intercalant, and Fig.~5\textbf{c} represents the combined contribution of the remaining three Nb sites (Sites 2, 3, 4) forming the kagome sublattice.

In the dashed box of Fig.~5\textbf{b}, the spectral weight of the Dirac crossing and the flat band (top) is markedly reduced for the central Nb site beneath Co, reflecting the strong hybridization with the Co $d_{z^2}$ orbital that quenches its contribution. On the other hand, the partially flat band below the Dirac crossing carries significant weight from the central Nb site. In contrast, Fig.~5\textbf{c} shows that the kagome Nb sites $d_{xy}$ orbital retains a much larger weight to the Dirac crossing and the flat band (top), underscoring their dominant role in stabilizing this kagome-like band feature near $E_F$. This clear site selectivity demonstrates how Co intercalation reconstructs the Nb sublattice, effectively suppressing the electronic participation of the central Nb atoms while enhancing that of the kagome sites. The structural origin of this effect is illustrated in Fig.~5\textbf{d}. The pristine hexagonal Nb network can be mapped onto an effective kagome lattice by excluding the central Nb site sitting directly below the Co site. The remaining three Nb atoms per unit cell define the frustrated kagome geometry (see the right panel of Fig.~5\textbf{d}), providing a natural framework for interpreting the low-energy states. This mapping emphasizes the evolution from a hexagonal layer to an effective kagome lattice induced by intercalation, which governs the emergence of flat-band physics in intercalated XY$_4$Z$_8$ compounds.

To further validate this mechanism, we constructed a tight-binding (TB) model (see Methods), the results of which are presented in Fig.~5\textbf{e}. In the case of the hexagonal lattice where all on-site potentials are equivalent (top panel), the four Nb atoms within the unit cell participate equally in the electronic states, leading to a set of strongly dispersive bands. This situation represents the unperturbed Nb network prior to the influence of intercalation. When the onsite potential of the central Nb site is suppressed to mimic the effect of Co site on top (middle panel), a band above the Dirac crossing flattens, consistent with the orbital-resolved DFT results. Notably, the effective kagome-like band structure highlighted in Fig.~5\textbf{c} matches closely with the kagome-like TB spectrum in the middle panel of Fig.~5\textbf{e}. The spectral weight of this flat band is thus governed predominantly by the three kagome-like Nb sites, while the central Nb site beneath the Co intercalant is quenched by vertical hybridization. In the limiting case of an ideal kagome lattice (bottom panel), this weakly dispersive band becomes perfectly flat, demonstrating the intrinsic kagome origin of the feature. Together, these results confirm that the flat-band state is governed predominantly by the effective kagome-like Nb sublattice with the suppressed central Nb sites. Similar kagome-derived flat-band features are also observed in the Co-intercalated TaSe$_2$ compound (see Supplementary Fig.~\textbf{S4}), highlighting that this mechanism is not unique to CoNb$_4$Se$_8$ but extends more broadly across the intercalated XY$_4$Z$_8$ family.

Together, these results demonstrate that the X intercalation reconstructs the Y sublattice of XY$_4$Z$_8$ into an effective kagome-like network by quenching the central Y $d_{z^2}$ orbital and enhancing the role of the kagome Y sites. The coexistence of weakly dispersive bands with Dirac-like dispersions at the K-point yields an orbital-selective electronic structure where localized kagome states and highly mobile Dirac carriers coexist in close energetic proximity. These features are particularly compelling for correlated electron physics, as the flat bands may host magnetically or electronically ordered phases, while the Dirac bands retain high mobility and topological character. Similar orbital-selective phenomena have been reported across a range of kagome systems, including CoSn, YMn$_6$Sn$_6$, FeSn, AV$_3$Sb$_5$, and, more recently, YbTi$_3$Bi$_4$ and Ti-based kagome lattices~\cite{Kang2020, Li2021, Shakya2023, Hu2023}, confirming the generality of this mechanism and establishing intercalated XY$_4$Z$_8$ compounds as a new platform within the broader kagome family.

In summary, our first-principles study of intercalated TMDs of the form XY$_4$Z$_8$ establishes a unified framework linking structural geometry, magnetic ordering, and orbital hybridization to emergent quantum phases. By correlating the interlayer-to-intralayer spacing ratio ($u/v$) with exchange pathways, we identified new A-type AFM compounds that realize nonrelativistic altermagnetism with momentum-dependent spin splitting, thereby extending the family of altermagnetic compounds. In the altermagnetic XY$_4$Z$_8$ compounds, the inclusion of SOC transforms the altermagnetic nodal lines into SOC-stabilized bulk Weyl points and surface Fermi arcs. Beyond magnetism and Weyl fermion topology, we demonstrated that intercalation drives the formation of kagome-like band structure by reconstructing the hexagonal Y sublattice, producing an orbital-selective landscape where localized (flat) and dispersive (Dirac) states coexist. These findings highlight intercalated XY$_4$Z$_8$ TMDs as a versatile platform for engineering altermagnetism, flat bands, and Weyl fermions within a single material family, offering general design principles for realizing correlated and topological quantum matter.

\newpage
\section*{Methods}
\subsection*{DFT Calculation}
First-principles calculations based on Density Functional Theory (DFT) were performed using the Vienna \textit{ab initio} Simulation Package (VASP). The exchange-correlation interactions were treated with the generalized gradient approximation (GGA) using the Perdew-Burke-Ernzerhof (PBE) functional. The core-valence electrons were described using the projector-augmented wave (PAW) method. A plane-wave kinetic energy cutoff of 450~eV, and a Monkhorst-Pack $9 \times 9 \times 4$ \textit{k}-point mesh were utilized in band structure calculations.

The simulated $2 \times 2 \times 1$ supercell consists of two intercalated transition-metal atoms positioned within the van der Waals (vdW) gaps between adjacent TMD layers, see Figs.~1\textbf{a} and 1\textbf{b}. The atomic positions and lattice parameters were optimized using the conjugate gradient method. All relaxations were performed by sampling both in-plane and out-of-plane FM and AFM spin alignments to determine the energetically favorable ground-state magnetic ordering.

To assess the robustness of the predicted antiferromagnetic ground states and account for on-site electron correlation effects in the partially filled transition-metal $d$ orbitals, additional calculations were performed within the GGA+$U$ framework using the Dudarev implementation. Effective on-site Coulomb interaction parameters were applied to the intercalated transition-metal and systematically varied over the range $U_{\mathrm{eff}} = 0$--3~eV. The results are shown in Supplementary Information.

\subsection*{Tight Binding Model}
To provide a qualitative understanding of weakly dispersive (flat) bands and the linear crossing points revealed by our \textit{ab initio} calculations, we construct a minimal tight-binding (TB) model that captures the essential physics of the Y sublattice in the intercalated XY$_4$Z$_8$ compounds. As illustrated in Fig.~5\textbf{d}, the Y atoms form a two-dimensional network that is modified by the presence of intercalants located directly above a subset of the Y sites. This structural motif naturally distinguishes one Y site, labeled $1$, which lies directly beneath an intercalant, from the three remaining Y sites ($2$, $3$, $4$) that are not directly aligned with intercalants. The resulting geometry interpolates between a four-site hexagonal lattice and a three-site kagome lattice, depending on the strength of the hopping to the site $1$ and on-site potential.

Within the nearest-neighbor approximation, the effective momentum-space Hamiltonian for the Y sublattice reads
\begin{equation}
\label{tb-h-kagome}
\resizebox{\textwidth}{!}{$
H_{HK} =
\begin{pmatrix}
\epsilon_{{\rm Y}_1} & -2 t'\cos{\left( {\kv}
 \cdot \av_1 \right)} & -2 t'\cos{\left({\kv}
 \cdot \av_2\right)} & -2 t'\cos{\left({\kv}
 \cdot \av_3\right)}\\
-2 t'\cos{\left({\kv}
 \cdot \av_1\right)} & \epsilon_{{\rm Y}_2} & -2 t\cos{\left({\kv}
 \cdot \av_3\right)} & -2 t\cos{\left({\kv}
 \cdot \av_2\right)}\\
-2 t'\cos{\left({\kv}
 \cdot \av_2\right)} & -2 t\cos{\left({\kv}
 \cdot \av_3\right)} & \epsilon_{{\rm Y}_3} & -2 t\cos{\left({\kv}
 \cdot \av_1\right)}\\
-2 t'\cos{\left({\kv}
 \cdot \av_3\right)} & -2 t\cos{\left({\kv}
 \cdot \av_2\right)} & -2 t\cos{\left({\kv}
 \cdot \av_1\right)} & \epsilon_{{\rm Y}_4}
\end{pmatrix},
$}
\end{equation}
where $\epsilon_{{\rm Y}_i}$ with $i=\overline{1,4}$ are the on-site potentials, $t$ is the hopping strength between the Y-sites that do not have X-sites on top of them ($i=2,3,4$), and $t'$ is the hopping strength to/from the atoms that have intercalants on top ($i=1$). The sublattice vectors $a_1, a_2, a_3$ define the geometry and connectivity of the model, and are defined as
$a_1=a\left\{1,0\right\}$, $a_2=a\left\{\tfrac{1}{2}, \tfrac{\sqrt{3}}{2}\right\}$, and $a_3=a\left\{-\tfrac{1}{2}, \tfrac{\sqrt{3}}{2}\right\}$ with $a$ being the inter-site distance. The Hamiltonian $H_{HK}$ describes the hexagonal lattice when all hopping constants and onsite potentials are equivalent, $t=t'$ and $\epsilon_{{\rm Y}_i}=\epsilon_{Y}$. In the limit $\epsilon_{{\rm Y}_1} \to \infty$ or $t'\to0$, the sites aligned with the intercalant become inaccessible, leading to a regular kagome lattice, see Fig.~5\textbf{d}. We show the evolution of the band spectrum with the relative contribution of each of the sites in Fig.~5\textbf{e}. This effective model bears a similarity with the \textit{ab initio} results for a smaller hopping strength $t'$ and an additional onsite potential $\epsilon_{{\rm Y}_1}$, as shown in Fig.~5\textbf{c} and the middle panel of Fig.~5\textbf{e}. This similarity between \textit{ab initio} and tight-binding spectra supports our interpretation of the origin of kagome-like features. The difference in the onsite potentials leads to the flattening of the upper and lower bands while preserving the linear crossing point ultimately resulting in an ideal kagome spectrum at $\epsilon_{{\rm Y}_1} \to \infty$ or $t'\to0$, as shown in the lower panel of Fig.~5\textbf{e}.

\newpage
\section*{Data availability}
All data needed to evaluate the conclusions in the paper are available within the article. All raw data generated during the current study are available from the corresponding author on reasonable request.

\section*{Acknowledgement}
The work at the University of Missouri, including the materials search and first-principles simulations, was primarily supported by the U.S. Department of Energy, Office of Science, Office of Basic Energy Sciences, Division of Materials Science and Engineering, under Grant No. DE-SC0024294. T.-R.C. was supported by National Science and Technology Council (NSTC) in Taiwan (Program No. NSTC 114-2628-M-006-005-MY3 and NSTC113-2124-M-006-009-MY3), National Cheng Kung University (NCKU), Taiwan, and National Center for Theoretical Sciences, Taiwan. This research was supported, in part, by the Higher Education Sprout Project, Ministry of Education to the Headquarters of University Advancement at NCKU. T.-R.C. thanks the National Center for High Performance Computing (NCHC) of National Applied Research Laboratories (NARLabs) in Taiwan for providing computational and storage resources. P.S. acknowledges useful communications with Onur Erten.

\section*{Author Information}
These authors contributed equally: Avinash Sah, Ting-Yong Lim. Correspondence and requests for materials should be addressed to Guang Bian (biang@missouri.edu), Tay-Rong Chang (u32trc00@phys.ncku.edu.tw), and Pavlo Sukhachov (pavlo.sukhachov@missouri.edu).

\subsection*{Contributions}
A.S., P.S., and G.B. conceived the idea. A.S., T.-Y.L., and T.-R.C. performed the DFT calculations. T.-Y.L., T.-R.C., P.S., and A.C. built the tight-binding model. P.S., G.B., C.C., and G.V. provided theoretical support. A.S., P.S., and G.B. wrote the manuscript with inputs from all authors. G.B., P.S., and T.-R.C. supervised the project. All the authors discussed the results.

\section*{Ethics declarations}
\subsection*{Competing interests}
The authors declare no competing financial or non-financial interests.

\newpage
\bibliographystyle{naturemag}
\bibliography{148}

\newpage
\begin{figure}
\includegraphics[width=1.0\linewidth]{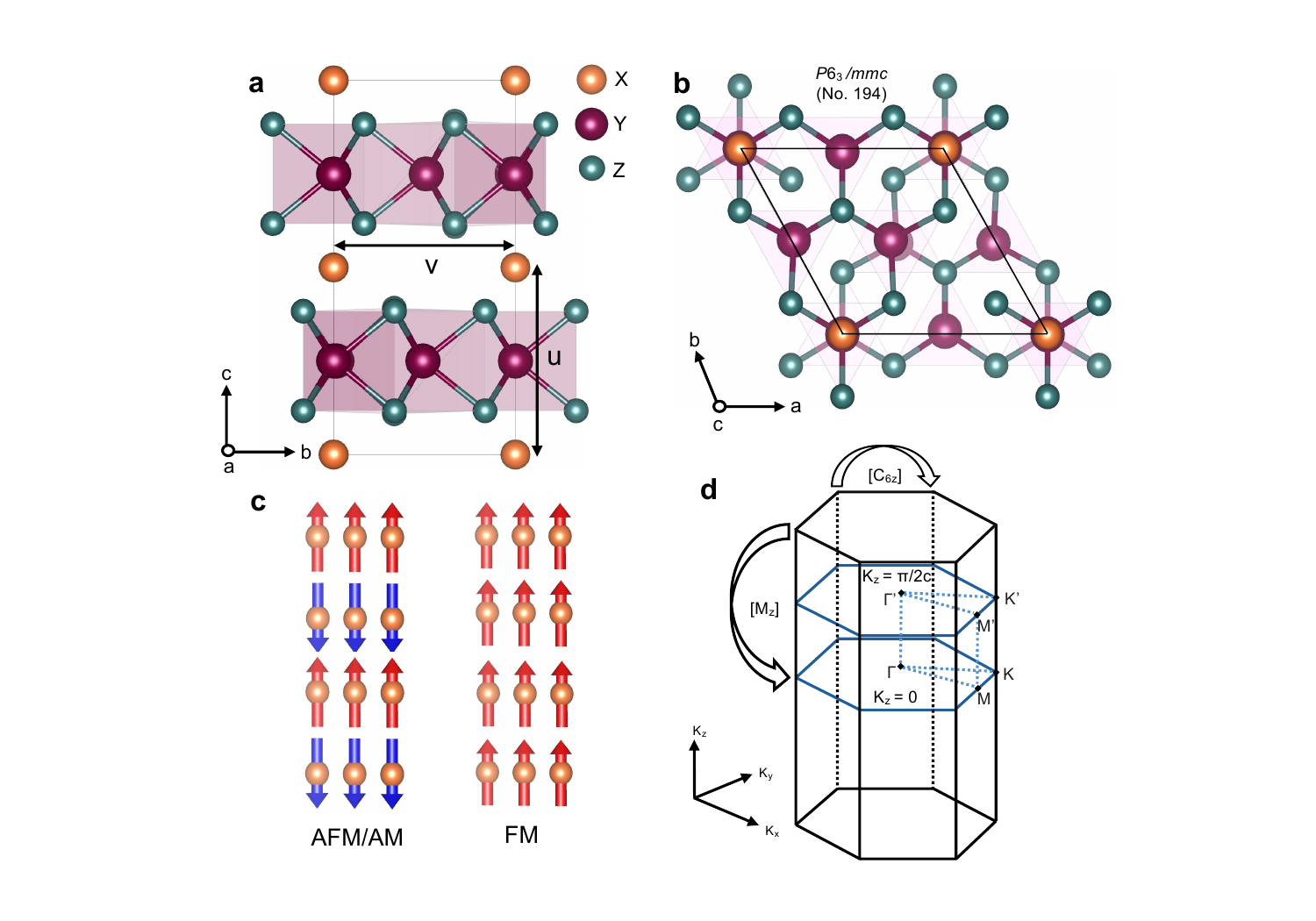}
\caption{\sffamily
\textbf{Crystal structure and magnetic configurations in intercalated transition metal dichalcogenides (TMDs).}
\textbf{a} Side view and \textbf{b} top view of the crystal structure, illustrating the atomic arrangement, where X represents intercalating transition metals (V, Cr, Mn, Fe, Co, Ni), Y represents host transition metals (Nb, Ta), and Z represents chalcogens (Se, S). The interlayer spacing ($\mathsf{u}$) and intralayer distance ($\mathsf{v}$) are indicated.
\textbf{c} Schematic illustration of magnetic configurations, depicting antiferromagnetic/altermagnetic (AFM/AM) and ferromagnetic (FM) alignments of spins in the layered structure. Arrows indicate spin directions.
\textbf{d} Three-dimensional Brillouin zone for the hexagonal lattice, highlighting symmetry points and paths used in band structure calculations. Spin-degenerate nodal planes are protected by the $\mathsf{C_{6z}}$ and $\mathsf{M_z}$ symmetries.
}
\end{figure}

\newpage
\begin{figure}
\includegraphics[width=1.0\linewidth]{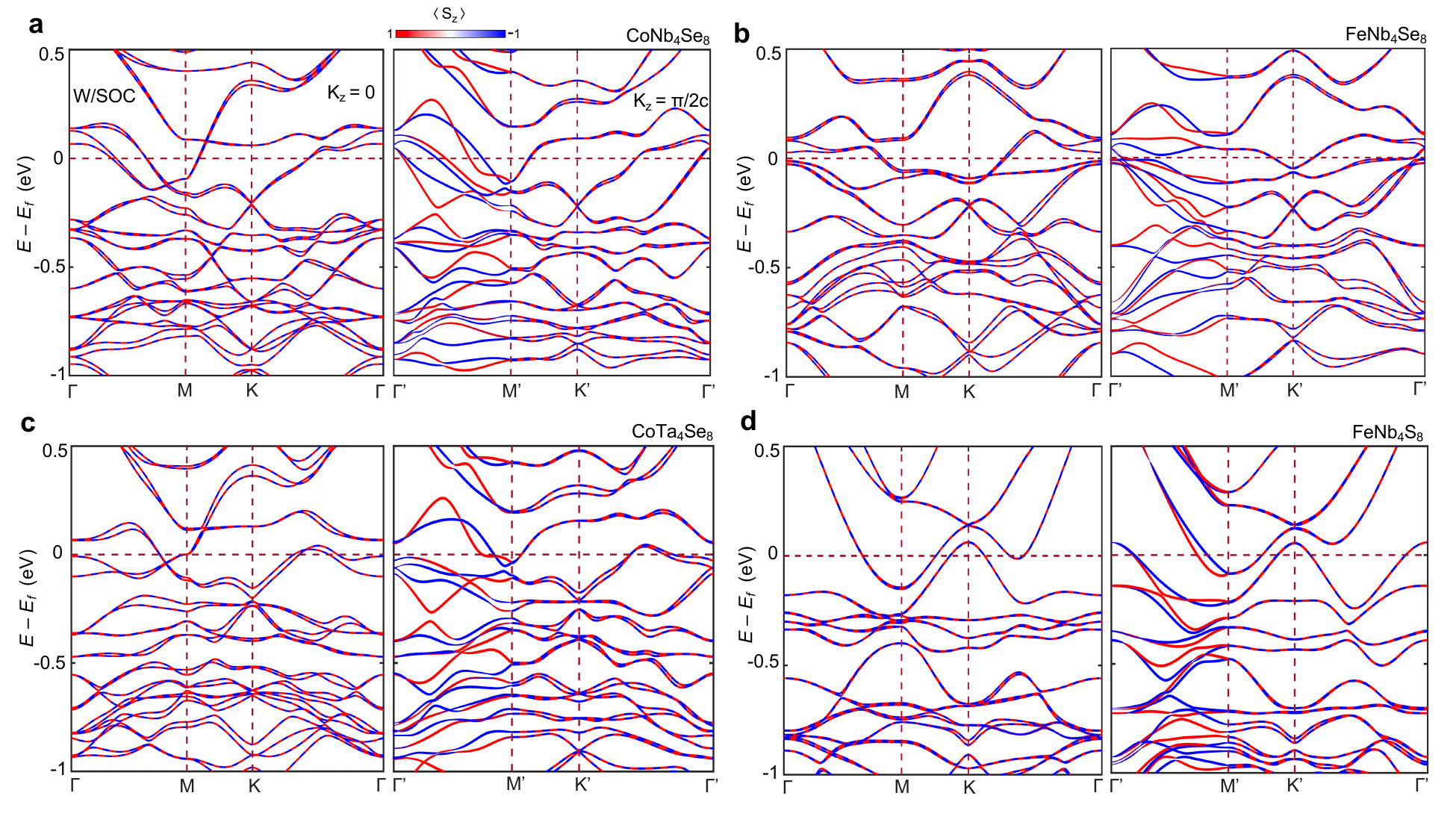}
\caption{\sffamily
\textbf{Calculated spin-resolved electronic band structures of altermagnetic intercalated TMDs with the inclusion of SOC.}
\textbf{a} CoNb$_{\mathsf{4}}$Se$_{\mathsf{8}}$, \textbf{b} FeNb$_{\mathsf{4}}$Se$_{\mathsf{8}}$, \textbf{c} CoTa$_{\mathsf{4}}$Se$_{\mathsf{8}}$, and \textbf{d} FeNb$_{\mathsf{4}}$S$_{\mathsf{8}}$.
For each compound, the band structures are shown on the (left) nodal plane ($\mathsf{k_z = 0}$) along the high-symmetry path $\mathsf{\Gamma}$–$\mathsf{M}$–$\mathsf{K}$–$\mathsf{\Gamma}$, and the (right) off-nodal plane ($\mathsf{k_z = \pi/2c}$) along $\mathsf{\Gamma^{\prime}}$–$\mathsf{M^{\prime}}$–$\mathsf{K^{\prime}}$–$\mathsf{\Gamma^{\prime}}$. Under SOC, the blue- and red-colored bands represent the spin projection $\langle S_z \rangle$ of SOC-mixed states, rather than pure spin-up and spin-down eigenstates.
Pronounced spin splitting is found along the off-nodal $\mathsf{\Gamma^{\prime}}$–$\mathsf{M^{\prime}}$ line.
}
\end{figure}

\newpage
\begin{figure}
\includegraphics[width=1.0\linewidth]{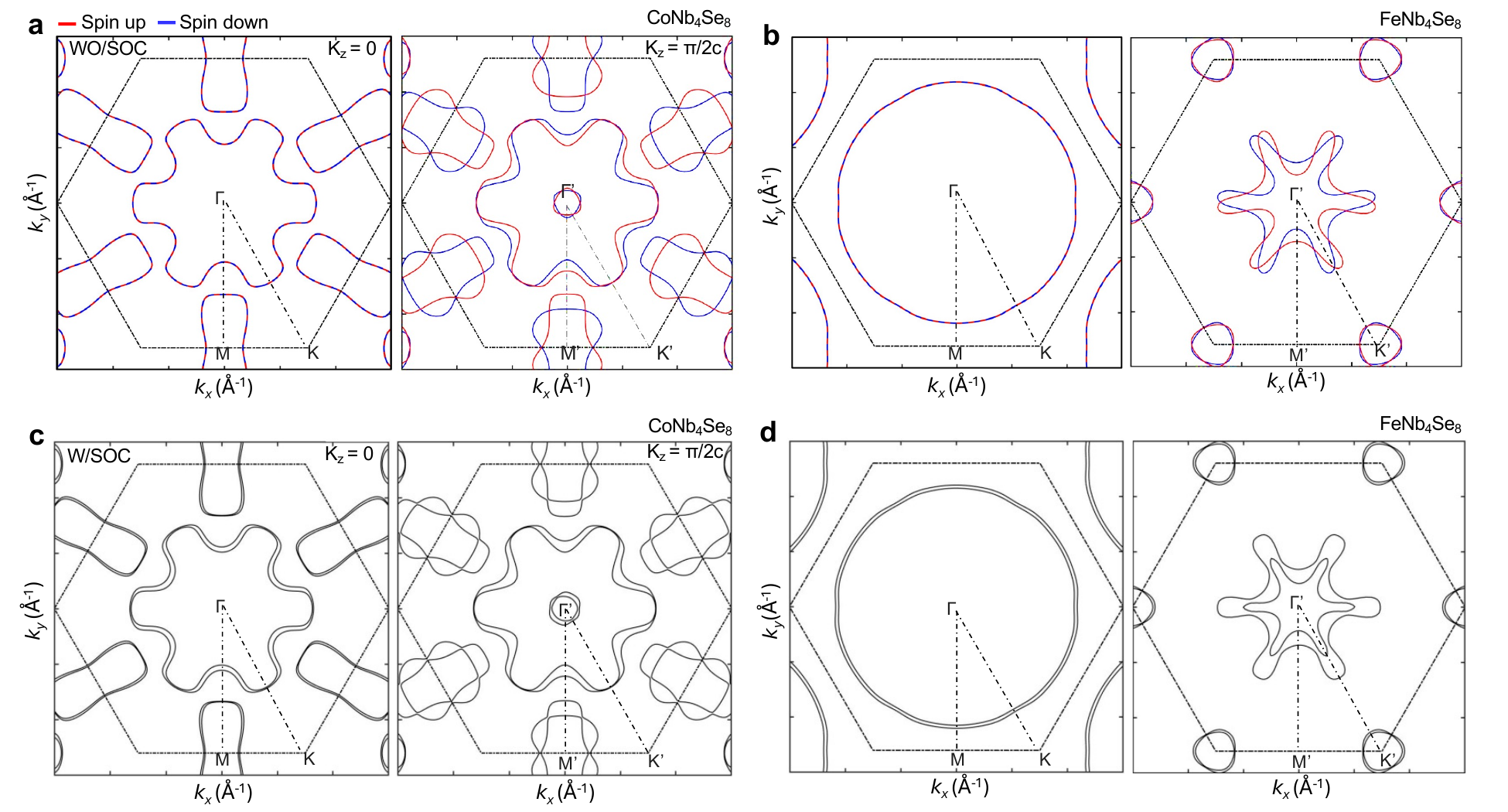}
\caption{\sffamily
\textbf{Calculated Fermi surface of altermagnetic intercalated TMDs.} \textbf{a} The spin-resolved Fermi surface of CoNb$_{\mathsf{4}}$Se$_{\mathsf{8}}$ in the absence of SOC at the planes with $\mathsf{k_{z}=0}$ (left) and $\mathsf{k_{z}=\pi/2c}$ (right). The red and blue contours correspond to spin-up and spin-down channels, respectively. \textbf{b} Same as \textbf{a}, but for FeNb$_{\mathsf{4}}$Se$_{\mathsf{8}}$. \textbf{c, d} The corresponding spectra with SOC included. All contours are plotted in black, as $s_z$ is no longer a good quantum number. Nevertheless, the residual altermagnetic spin texture closely resembles that without SOC.
%%as momentum-dependent band splittings that reconstruct the Fermi surface topology across the off-nodal plane along $\mathsf{k_{z}}$.
}
\end{figure}

\newpage
\begin{figure}
\includegraphics[width=1.0\linewidth]{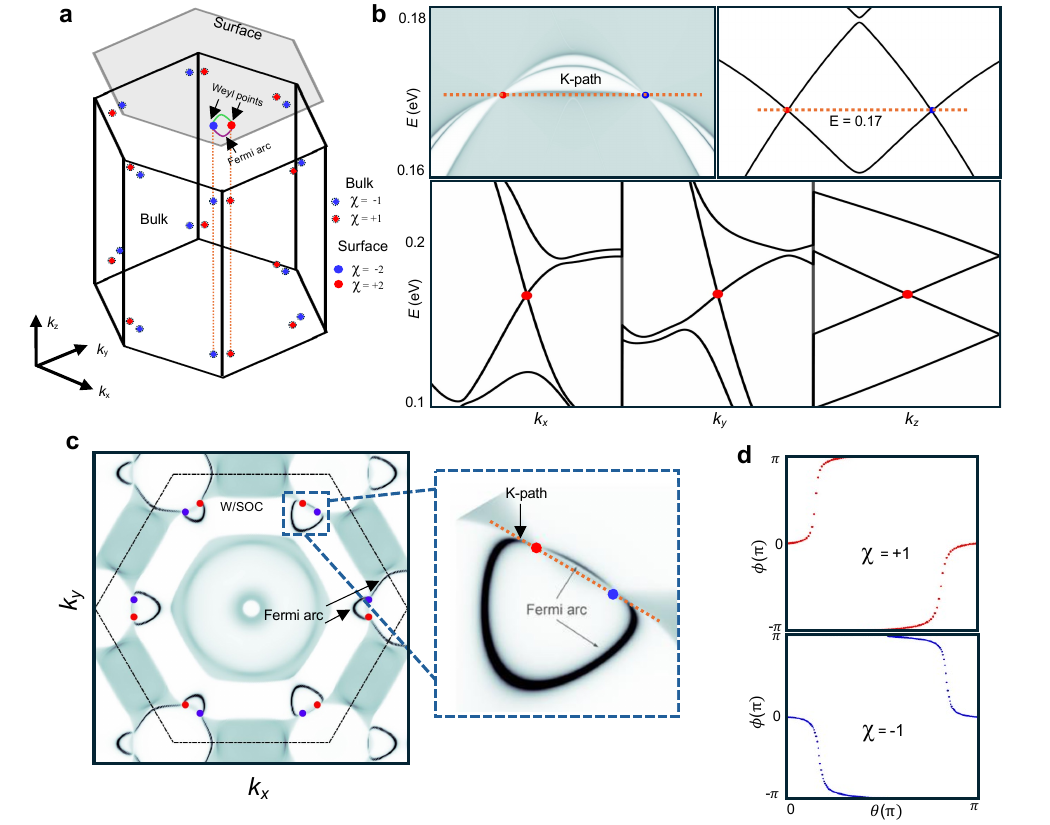}
\caption{\sffamily
\textbf{Weyl points and topological surface states in CoNb$_{\mathsf{4}}$Se$_{\mathsf{8}}$.}
\textbf{a} Distribution of WPs in the bulk Brillouin zone (BZ) and their projections onto the (001) surface BZs. Bulk WPs with chirality $\mathsf{\chi = \pm 1}$ are indicated in dashed red and blue. In cases where two same-chirality WPs are projected onto the same point in the surface Brillouin zone, the projections appear to have effective topological charges $\mathsf{\chi = \pm 2}$.
\textbf{b} Top left: zoom of the surface spectrum along the $\mathsf{K}$-path with the SOC-induced Weyl crossings. Top right: band structure showing band crossings along the same path with WPs located near $\mathsf{E = 0.17~eV}$. Bottom: band dispersions along the $\mathsf{k_x}$, $\mathsf{k_y}$, and $\mathsf{k_z}$ directions.
\textbf{c} Calculated surface spectrum of the (001) surface with SOC. The inset highlights the surface Fermi arcs (SFAs), where overlapping projections of two bulk Weyl nodes give rise to double Fermi arcs (black arrows).
\textbf{d} Evolution of spin-resolved hybrid Wannier charge centers around bulk WPs, with topological charges $\mathsf{\chi = +1}$ and $\mathsf{\chi = -1}$, demonstrating opposite chirality for isolated bulk WPs and nontrivial Berry curvature.
}
\end{figure}

\newpage
\begin{figure}
\includegraphics[width=1.0\linewidth]{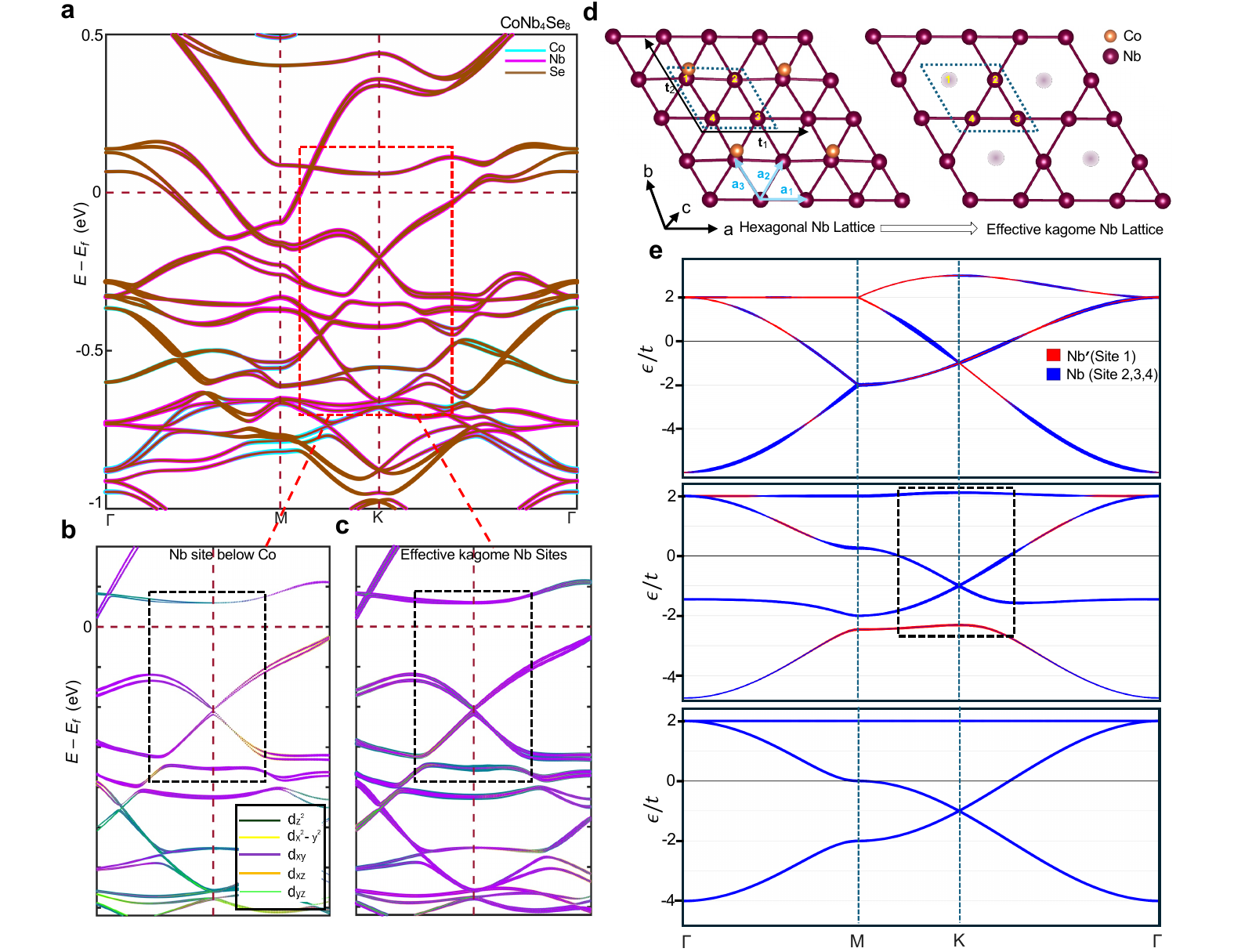}
\caption{\sffamily
\textbf{Kagome-like electronic and lattice structure in CoNb$_{\mathsf{4}}$Se$_{\mathsf{8}}$.}
\textbf{a} Orbital-projected band structure highlighting contributions from Co, Nb, and Se.
\textbf{b, c} Orbital-resolved Nb contributions comparing the site directly beneath the Co intercalant (left) with the kagome-like Nb sites (right).
\textbf{d} Top view of the pristine hexagonal Nb network (left) can be mapped onto an effective kagome lattice (right) by removing the Nb site directly below each Co atom (Site 1). The remaining Nb sites (Sites 2, 3, and 4) define the kagome sublattice. Numbered Nb sites denote sites within the reduced lattice, and the dashed box corresponds to the unit cell chosen for the TB model construction. Primitive lattice vectors $\mathsf{t_{1}}$, $\mathsf{t_{2}}$ and sublattice vectors $\mathsf{a_{1}}$, $\mathsf{a_{2}}$, $\mathsf{a_{3}}$ define the geometry and connectivity of the model.
\textbf{e} Energy spectra of the model in Eq.~\eqref{tb-h-kagome} for three Nb lattices: hexagonal (top), kagome-like (middle), and ideal kagome (bottom). Nb$^{\prime}$ (red) denotes the site directly beneath Co, and Nb (blue) denotes the kagome sites. In the top panel (hexagonal), all hoppings and onsite potentials are equal. In the middle panel (kagome-like), we set $\mathsf{t'/t}=-0.4$ and $\mathsf{\epsilon_{Nb'}/t}=-2.2$ (all other onsite terms zero). In the bottom panel (ideal kagome), we take $\mathsf{t'/t}=0$ and $\mathsf{\epsilon_{Nb'}/t}=-100$. Parameters in the middle panel are chosen to resemble the DFT band structure in Fig.~5\textbf{c}.
}
\end{figure}

%\begin{comment}

\begin{table}[h!]
\centering
\renewcommand{\arraystretch}{1.1}
\setlength{\tabcolsep}{4pt}
\begin{threeparttable}
%\resizebox{\textwidth}{!}{
\begin{adjustbox}{max width=\textwidth}
\begin{tabular}{ccccc}
\toprule
\textbf{XY$_4$Z$_8$} & \textbf{Interlayer($u$,\AA)} & \textbf{Intralayer($v$,\AA)} & \textbf{Ratio ($u/v$)} & \textbf{GS} \\
\midrule
FeNb$_4$S$_8$  & 5.838 & 6.656 & 0.877 & AFM/AM \\
CoNb$_4$Se$_8$ & 6.232 & 6.978 & 0.893 & AFM/AM \\
NiNb$_4$Se$_8$ & 6.266 & 6.989 & 0.897 & FM \\
CoTa$_4$Se$_8$ & 6.281 & 6.930 & 0.906 & AFM/AM \\
FeNb$_4$Se$_8$ & 6.354 & 6.987 & 0.909 & AFM/AM \\
CrNb$_4$Se$_8$ & 6.364 & 6.983 & 0.911 & FM \\
VNb$_4$Se$_8$  & 6.421 & 6.986 & 0.919 & FM \\
MnNb$_4$Se$_8$ & 6.505 & 7.009 & 0.928 & FM \\
MnTa$_4$S$_8$  & 6.238 & 6.672 & 0.935 & FM \\
\bottomrule
\end{tabular}
%}
\end{adjustbox}
\caption{\sffamily
\textbf{Structural parameters of intercalated TMDs (XY$_{\mathsf{4}}$Z$_{\mathsf{8}}$).}
X = Mn, Fe, Co, Ni, Cr, V; Y = Nb, Ta; Z = Se, S.
Shown are the interlayer ($\mathsf{u}$) spacing, intralayer ($\mathsf{v}$) spacing between intercalated atoms, and their ratio $\mathsf{u/v}$.
The corresponding magnetic ground states (GS), obtained from DFT calculations, are also indicated: FM = ferromagnetic; AFM/AM = A-type antiferromagnetic/altermagnetic.
}
\label{Table1}
\end{threeparttable}
\end{table}
%\end{comment}

\begin{comment}
\newcolumntype{Y}{>{\centering\arraybackslash}X}

\begin{table}[h!]
\centering
\renewcommand{\arraystretch}{1.1}
\setlength{\tabcolsep}{4pt}
\begin{threeparttable}
\begin{tabularx}{\textwidth}{Y Y Y Y Y}
%\resizebox{\textwidth}{!}{
%\begin{adjustbox}{max width=\textwidth}
%\begin{tabular}{ccccc}
\toprule
\textbf{XY$_4$Z$_8$} & \textbf{Interlayer($u$,\AA)} & \textbf{Intralayer($v$,\AA)} & \textbf{Ratio ($u/v$)} & \textbf{GS} \\
\midrule
FeNb$_4$S$_8$  & 5.838 & 6.656 & 0.877 & AFM/AM \\
CoNb$_4$Se$_8$ & 6.232 & 6.978 & 0.893 & AFM/AM \\
NiNb$_4$Se$_8$ & 6.266 & 6.989 & 0.897 & FM \\
CoTa$_4$Se$_8$ & 6.281 & 6.930 & 0.906 & AFM/AM \\
FeNb$_4$Se$_8$ & 6.354 & 6.987 & 0.909 & AFM/AM \\
CrNb$_4$Se$_8$ & 6.364 & 6.983 & 0.911 & FM \\
VNb$_4$Se$_8$  & 6.421 & 6.986 & 0.919 & FM \\
MnNb$_4$Se$_8$ & 6.505 & 7.009 & 0.928 & FM \\
MnTa$_4$S$_8$  & 6.238 & 6.672 & 0.935 & FM \\
\bottomrule
\end{tabularx}
%\end{tabular}
%}
\end{adjustbox}
\caption{\sffamily
\textbf{Structural parameters of intercalated TMDs (XY$_{\mathsf{4}}$Z$_{\mathsf{8}}$).}
X = Mn, Fe, Co, Ni, Cr, V; Y = Nb, Ta; Z = Se, S.
Shown are the interlayer ($\mathsf{u}$) spacing, intralayer ($\mathsf{v}$) spacing between intercalated atoms, and their ratio $\mathsf{u/v}$.
The corresponding magnetic ground states (GS), obtained from DFT calculations, are also indicated: FM = ferromagnetic; AFM/AM = A-type antiferromagnetic/altermagnetic.
}
\label{Table1}
\end{threeparttable}
\end{table}
\end{comment}

\end{document}

\title{Supplementary Information \\ Altermagnetism, Kagome Flat Band, and Weyl Fermion States in Magnetically Intercalated Transition Metal Dichalcogenides}

\author{Avinash Sah}
\affiliation {Department of Physics and Astronomy, University of Missouri, Columbia, Missouri 65211, USA}

\author{Ting-Yong Lim }
\affiliation {Department of Physics, National Cheng Kung University, Tainan 701, Taiwan}

\author{Clayton Conner}
\affiliation {Department of Physics and Astronomy, University of Missouri, Columbia, Missouri 65211, USA}

\author{Amarnath Chakraborty}
\affiliation {Department of Physics and Astronomy, University of Missouri, Columbia, Missouri 65211, USA}

\author{Giovanni Vignale}
\affiliation {Department of Physics and Astronomy, University of Missouri, Columbia, Missouri 65211, USA}

\author{Tay-Rong Chang}\email{u32trc00@phys.ncku.edu.tw}
\affiliation {Department of Physics, National Cheng Kung University, Tainan 701, Taiwan}
\affiliation {Center for Quantum Frontiers of Research and Technology (QFort), Tainan, 70101, Taiwan}
\affiliation {Physics Division, National Center for Theoretical Sciences, Taipei, 10617, Taiwan}

\author{Pavlo Sukhachov}\email{pavlo.sukhachov@missouri.edu}
\affiliation {Department of Physics and Astronomy, University of Missouri, Columbia, Missouri 65211, USA}
\affiliation {MU Materials Science \& Engineering Institute, University of Missouri, Columbia, Missouri 65211, USA}

\author{Guang~Bian}\email{biang@missouri.edu}
\affiliation {Department of Physics and Astronomy, University of Missouri, Columbia, Missouri 65211, USA}
\affiliation {MU Materials Science \& Engineering Institute, University of Missouri, Columbia, Missouri 65211, USA}

\maketitle
\newpage

% =========================
% ====== FIGURE S1 ========
% =========================
\begin{figure}
\includegraphics[width=1.0\linewidth]{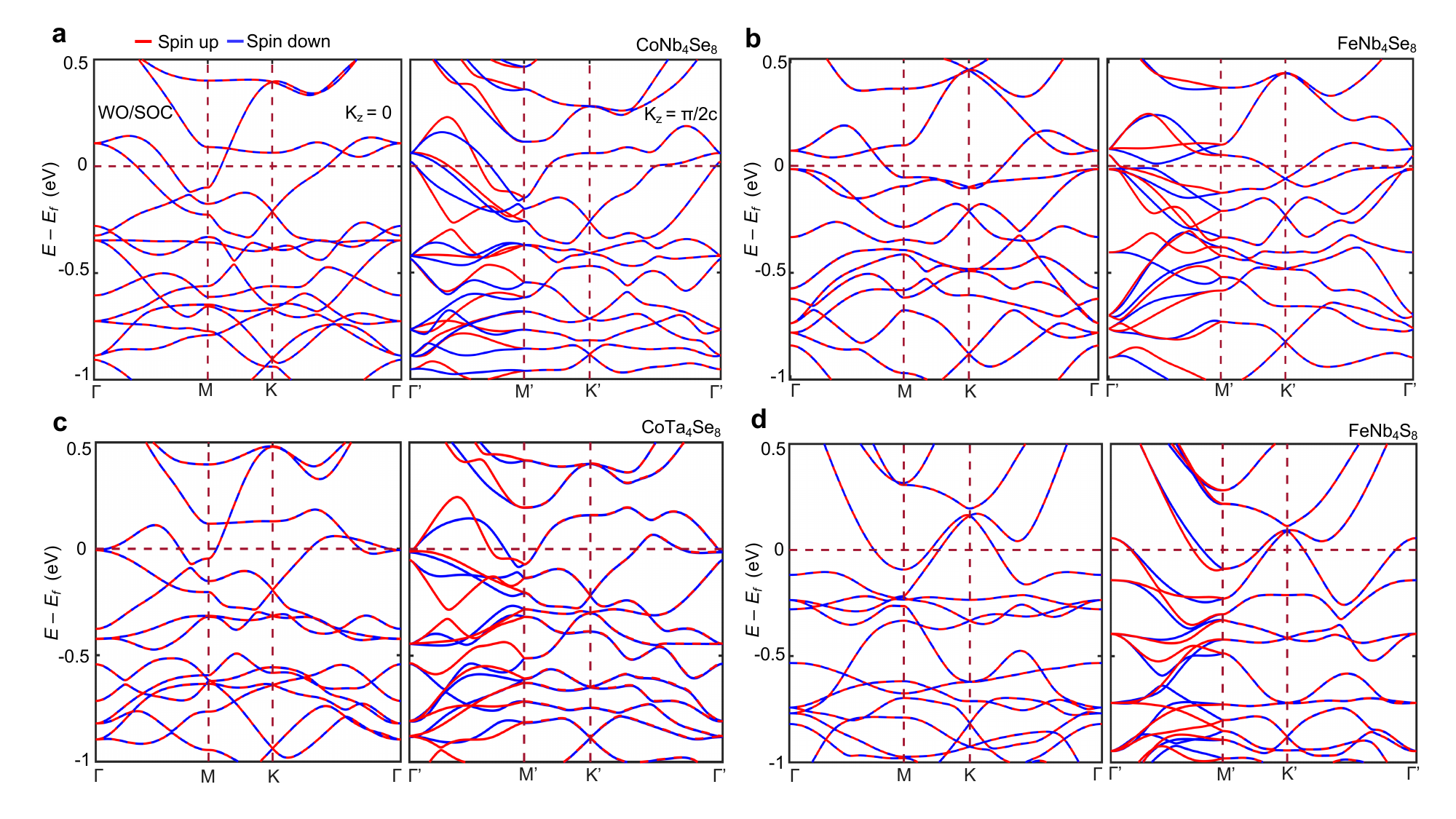}
\caption{\sffamily
\textbf{DFT-calculated spin-resolved electronic band structures of altermagnetic intercalated TMDs without SOC.}
\textbf{a} CoNb$_{\mathsf{4}}$Se$_{\mathsf{8}}$, \textbf{b} FeNb$_{\mathsf{4}}$Se$_{\mathsf{8}}$, \textbf{c} CoTa$_{\mathsf{4}}$Se$_{\mathsf{8}}$, and \textbf{d} FeNb$_{\mathsf{4}}$S$_{\mathsf{8}}$.
For each compound, the band structures are shown on the (left) nodal plane ($\mathsf{k_z = 0}$) along the high-symmetry path $\mathsf{\Gamma}$–$\mathsf{M}$–$\mathsf{K}$–$\mathsf{\Gamma}$, and the (right) off-nodal plane ($\mathsf{k_z = \pi/2c}$) along $\mathsf{\Gamma^{\prime}}$–$\mathsf{M^{\prime}}$–$\mathsf{K^{\prime}}$–$\mathsf{\Gamma^{\prime}}$.
Red and blue lines represent spin-up and spin-down channels, respectively.
Pronounced spin splitting is observed along the off-nodal $\mathsf{\Gamma^{\prime}}$–$\mathsf{M^{\prime}}$ line.
}
\label{fig:S1}
\end{figure}

% =========================
% ====== FIGURE S2 ========
% =========================
\newpage
\begin{figure}
\centering
\includegraphics[width=\linewidth]{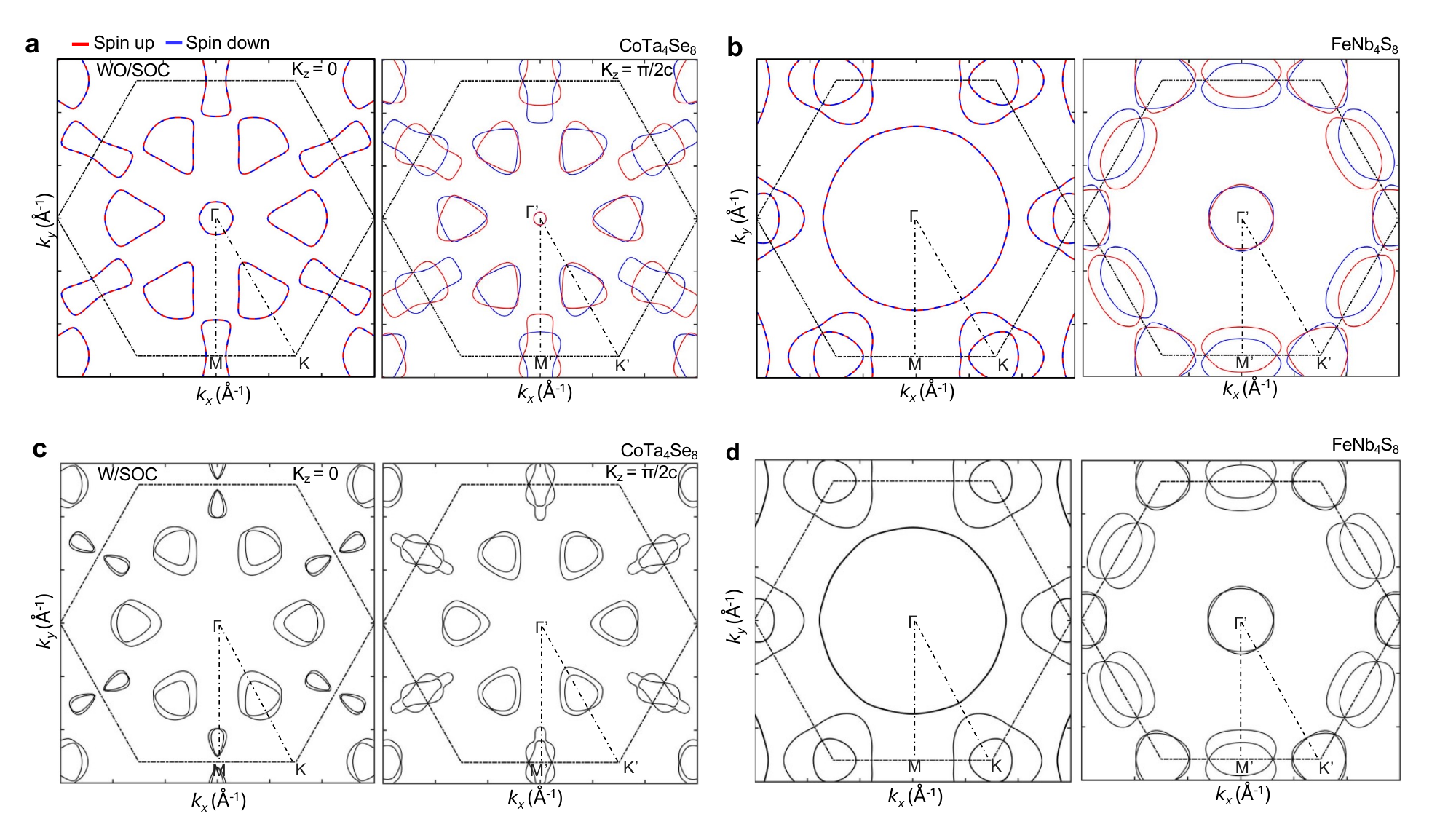}
\caption{\sffamily
\textbf{Fermi surface cross-sections of altermagnetic intercalated TMDs with and without SOC.}
Constant-energy contours are shown at the nodal plane ($\mathsf{k_{z}=0}$, left) and the off-nodal plane ($\mathsf{k_{z}=\pi/2c}$, right) for \textbf{a, c} CoTa$_{\mathsf{4}}$Se$_{\mathsf{8}}$ and \textbf{b, d} FeNb$_{\mathsf{4}}$S$_{\mathsf{8}}$.
Panels \textbf{a, b} display the spin-resolved Fermi surfaces in the absence of SOC, where red and blue contours correspond to spin-up and spin-down channels, respectively.
Panels \textbf{c, d} show the corresponding spectra with SOC included, for which spin is no longer a good quantum number, and all contours are plotted in black.
}
\label{fig:S2}
\end{figure}

% =========================
% ====== FIGURE S3 ========
% =========================
\newpage
\begin{figure}
\centering
\includegraphics[width=\linewidth]{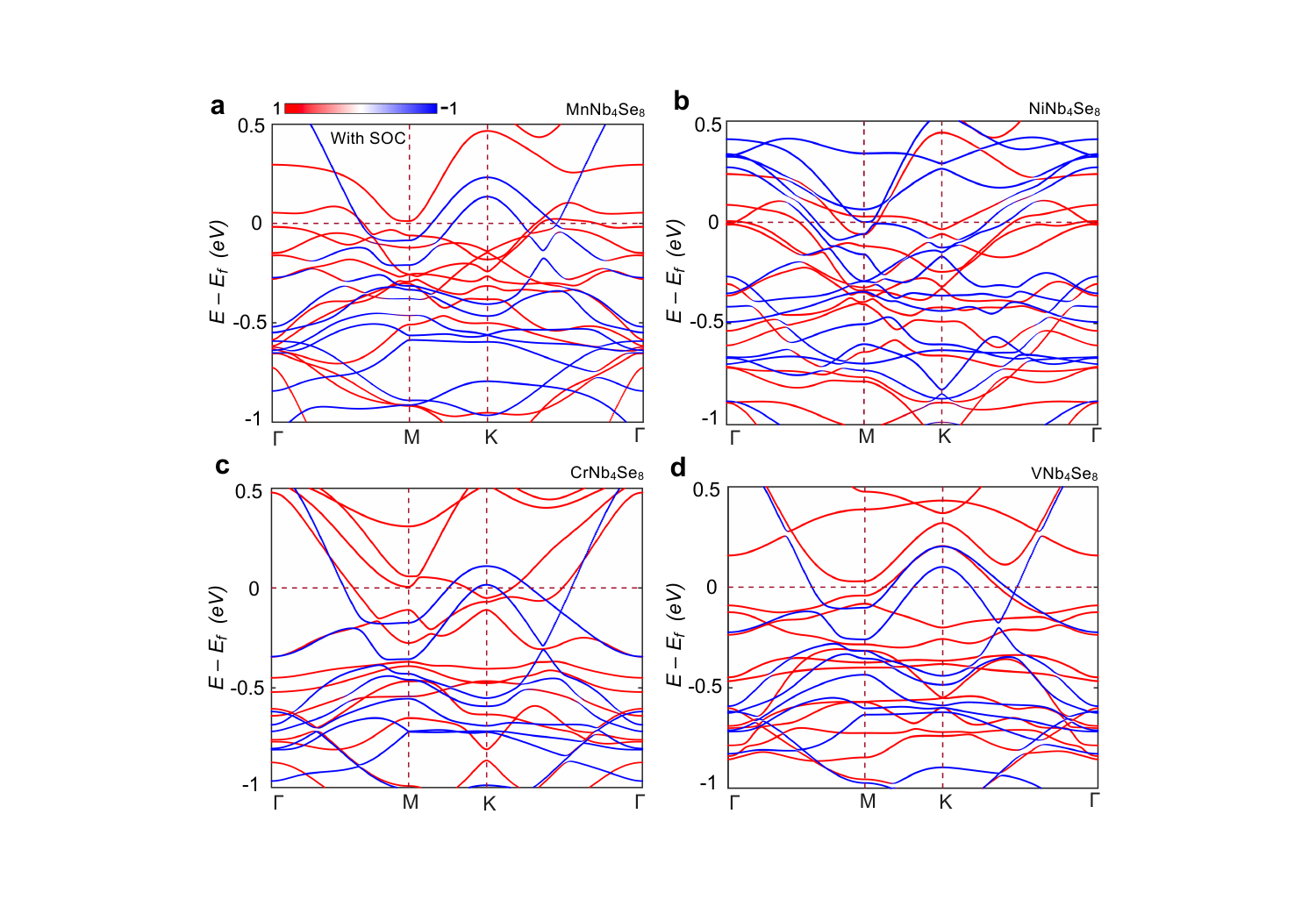}
\caption{\sffamily
\textbf{DFT-calculated spin-resolved electronic band structures of ferromagnetic intercalated TMDs.}
\textbf{a} MnNb$_{\mathsf{4}}$Se$_{\mathsf{8}}$, \textbf{b} NiNb$_{\mathsf{4}}$Se$_{\mathsf{8}}$, \textbf{c} CrNb$_{\mathsf{4}}$Se$_{\mathsf{8}}$, and \textbf{d} VNb$_{\mathsf{4}}$Se$_{\mathsf{8}}$.
Band structures are plotted along the high-symmetry path $\mathsf{\Gamma}$–$\mathsf{M}$–$\mathsf{K}$–$\mathsf{\Gamma}$ on the nodal plane ($\mathsf{k_z = 0}$).
}
\label{fig:S3}
\end{figure}

% =========================
% ====== FIGURE S4 ========
% =========================
\newpage
\begin{figure}
\centering
\includegraphics[width=\linewidth]{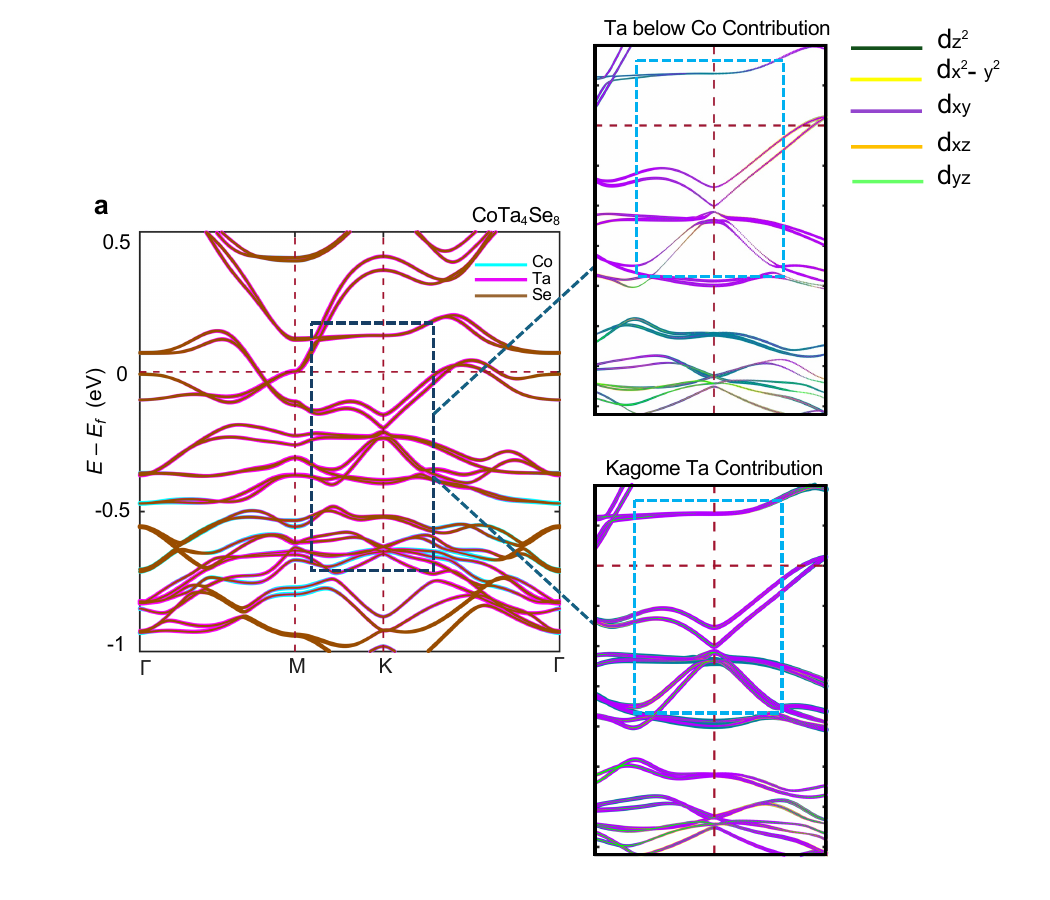}
\caption{\sffamily
\textbf{Atomic and orbital-resolved electronic band structures of CoTa$_{\mathsf{4}}$Se$_{\mathsf{8}}$.}
Top inset: orbital-resolved band structure of the Ta site located directly beneath Co.
Bottom inset: corresponding orbital-resolved band structure for the kagome Ta sites.
The dashed box marks the region near the Fermi level to show orbital contributions to the flat bands and Dirac-like dispersions.
}
\label{fig:S4}
\end{figure}

% =========================
% ====== FIGURE S5 ========
% =========================
\newpage
\begin{figure}
\centering
\includegraphics[width=\linewidth]{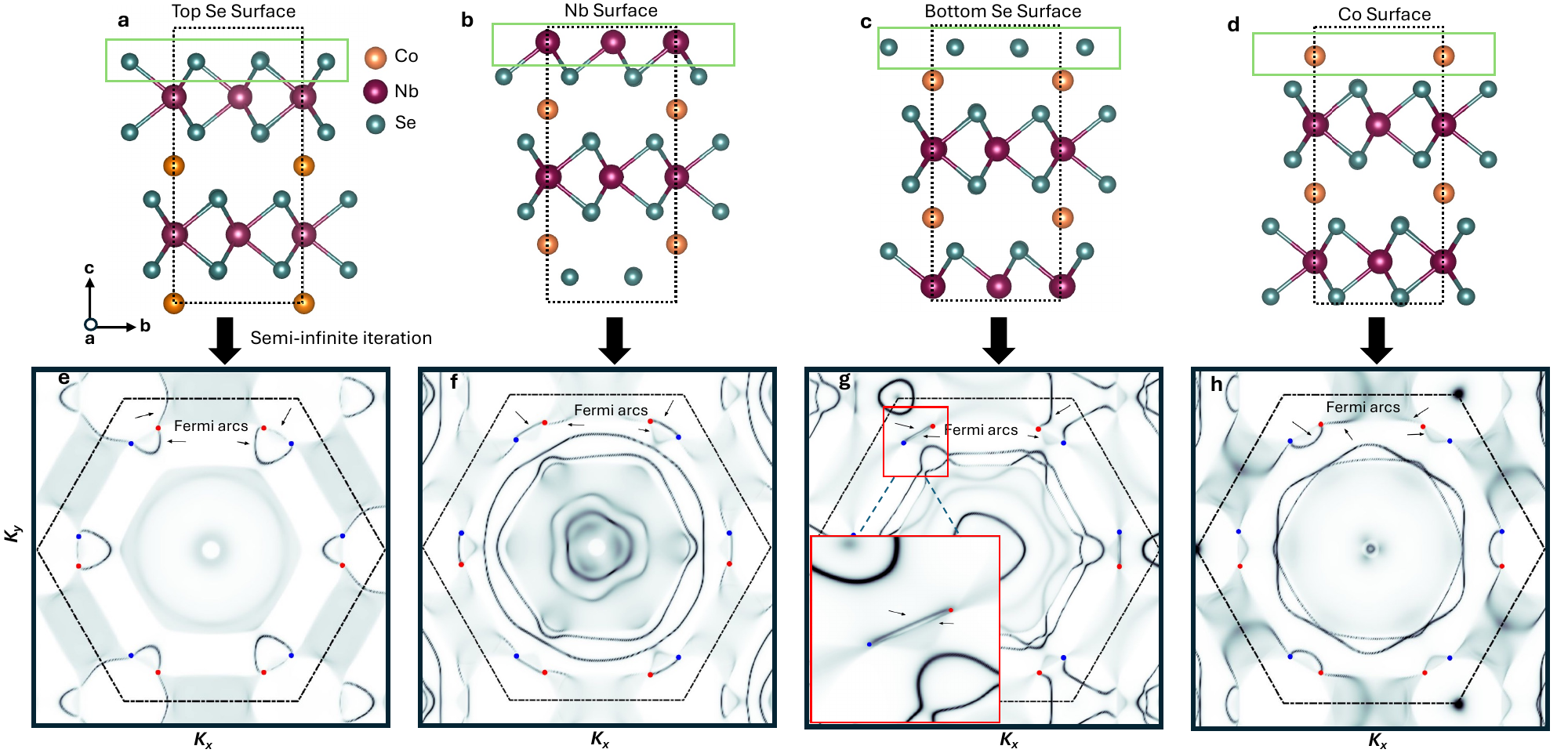}
\caption{\sffamily
\textbf{Robustness of Fermi arcs on different surface terminations of CoNb$_{\mathsf{4}}$Se$_{\mathsf{8}}$.}
(a–d) Schematic structures of four possible surface terminations used in the semi-infinite iteration calculations: (a) top Se-terminated, (b) Nb-terminated, (c) bottom Se-terminated, and (d) Co-terminated surfaces. The green box highlights the terminating layer. (e–h) Corresponding surface spectrum of different terminations projected onto the surface Brillouin zone. Red and blue dots mark the projected Weyl points of opposite chirality, while arrows indicate the Fermi-arc connecting them. Although the Weyl point projections remain fixed by bulk topology, the shape, curvature, and spectral distribution of the Fermi arcs vary with surface termination due to changes in the surface potential and orbital environment.
}
\label{fig:S5}
\end{figure}

\renewcommand{\thetable}{T\arabic{table}}
\setcounter{table}{0}

\begin{table}[t]
\centering
\makebox[\textwidth][c]{%
\begin{tabular}{lcc}
\hline
\textbf{Material}
& $\Delta_{\mathrm{spin}}$ at Fermi level (meV) 
& $\Delta_{\mathrm{spin}}$ at $-0.25$ eV (meV) \\
\hline
CoTa$_4$Se$_8$ & 110 & 154 \\
CoNb$_4$Se$_8$ &  84 & 128 \\
FeNb$_4$Se$_8$ &  76 & 135 \\
FeNb$_4$S$_8$  &  55 &  40 \\
\hline
\end{tabular}}
\caption{\sffamily
\textbf{Spin-splitting magnitude $\Delta_{\mathrm{spin}}$ evaluated at the Fermi level and at $\mathsf{-0.25}$ eV for representative magnetically intercalated transition metal dichalcogenides.}}
\label{tab:spin_splitting}
\end{table}

\begin{table}[t]
\centering
\makebox[\textwidth][c]{%
\begin{tabular}{lccccc}
\hline
\textbf{Material} &
\multicolumn{5}{c}{\boldsymbol{$\Delta E = E_{\mathrm{FM}} - E_{\mathrm{A\text{-}AFM}}$} (meV)} \\
\cline{2-6}
 & $U=0$ & $U=1$ & $U=2$ & $U=3$ \\
\hline
CoTa$_4$Se$_8$ & 199.5 & 448.7 & 584.3 & 663.1 \\
CoNb$_4$Se$_8$ & 120.6  & 419.9 & 652.2  & 834.38  \\
FeNb$_4$Se$_8$ & 62.7  & 96.2  & 183.1  & 261.4  \\
FeNb$_4$S$_8$  & 51.4 & 79.01  & 119.9 & 555.8 \\
\hline
\end{tabular}}
\caption{\sffamily
\textbf{Robustness of the magnetic ground state against onsite Coulomb interaction $U$.}
Energy differences $\Delta E = E_{\mathrm{FM}} - E_{\mathrm{A\text{-}AFM}}$ obtained from GGA+$U$ calculations for representative magnetically intercalated transition metal dichalcogenides, using the Dudarev approach.
Positive values indicate stabilization of the A-type antiferromagnetic (A-AFM) ground state over the full interaction range $U = 0$-3 eV, demonstrating that the magnetic ordering remains robust against electronic correlation effects.}
\label{tab:AFM_FM_energy}
\end{table}

\begin{table*}[t]
\centering
\renewcommand{\arraystretch}{1.25}
\setlength{\tabcolsep}{8pt}
\begin{tabular}{ll}
\hline
\multicolumn{2}{c}{\textbf{Symmetry Classification of altermagnetic XY$_4$Z$_8$}} \\
\hline
\textbf{Spin Space Group (SSG)} &
$P^{-1}6_3/^{-1}m^{1}m^{-1}c^{\infty m}1$ \\

\textbf{GO Symbol (Parent Space Group)} &
$P6_3/mmc\,(194)$ \\

\textbf{LO Symbol (Little Group of the Order Parameter)} &
$P\!-\!3m1\,(164)$ \\

\textbf{Magnetic Space Group (MSG)} &
$P6_3'/m'm'c\,(194.268)$ \; (Type III) \\

\textbf{Spin Part Point Group (Spin PG)} &
$C_i \times C_{\infty v} = D_{\infty h}$ \\
& $-1 \times \infty m = \infty/mm$ \\

\textbf{Effective Magnetic Point Group} &
$6/mmm1'$ \\

\hline
\end{tabular}
\caption{\sffamily
\textbf{Symmetry classification of altermagnetic XY$_4$Z$_8$.}
The spin space group (SSG) describes the full symmetry of the magnetic structure in the absence of spin-orbit coupling, where spin and spatial rotations are decoupled.
The GO symbol denotes the parent crystallographic space group of the nonmagnetic phase, while the LO symbol identifies the little group associated with the magnetic order parameter. The magnetic space group (MSG) is of Type III, indicating that pure time-reversal symmetry is broken but combined antiunitary symmetries are preserved. The spin part point group specifies the symmetry of the spin degrees of freedom, and the effective magnetic point group governs macroscopic response properties. This symmetry setting corresponds to an A-type antiferromagnetic altermagnet, allowing momentum-dependent spin splitting without net magnetization.
}
\end{table*}

\begin{table*}[tbp]
\centering
\renewcommand{\arraystretch}{1.2}
\setlength{\tabcolsep}{6pt}

\begin{tabular}{lcccc}
\hline
\textbf{Strain}
& $\Delta E_{\mathrm{A\!-\!AFM}}$ (meV)
& $\Delta E_{\mathrm{FM}}$ (meV)
& $\Delta E$ (meV)
& $\Delta E_{\mathrm{rel}} = \Delta E - \Delta E_{\mathrm{base}}$ \\
\hline
Base $(u_0,v_0)$ & 0 & 0 & +11.19 & 0.00 \\
$u=-1\%$ & 21.83 & 25.18 & +7.83 & \textbf{-3.36} \\
$u=-2\%$ & 75.62 & 84.04 & +2.78 & \textbf{-8.41} \\
$u=-3\%$ & 169.32 & 180.15 & +0.36 & \textbf{-10.83} \\
$u=+1\%$ & 18.70 & 2.41 & +27.48 & \textbf{+16.29} \\
$u=+2\%$ & 61.67 & 31.92 & +40.95 & \textbf{+29.76} \\
$u=+3\%$ & 126.80 & 85.40 & +52.60 & \textbf{+41.41} \\

\hline
\end{tabular}

\caption{\sffamily
\textbf{Evolution of magnetic ground-state energetics in NiNb$_4$Se$_8$ under strain.}
The first column lists the applied out-of-plane strain $u$ (equivalently a $c$-axis distortion) relative to the optimized base structure $(u_0, v_0)$. 
$\Delta E_{\mathrm{A\text{-}AFM}}$ and $\Delta E_{\mathrm{FM}}$ denote the strain-induced total-energy changes (meV) of the A-type antiferromagnetic (A-AFM) and ferromagnetic (FM) configurations, respectively, normalized to their ground-state minima at $(u_0, v_0)$. 
$\Delta E = E_{\mathrm{A\text{-}AFM}} - E_{\mathrm{FM}}$ is the magnetic energy difference at a given strain; positive values indicate stabilization of the FM state. 
$\Delta E_{\mathrm{rel}} = \Delta E - \Delta E_{\mathrm{base}}$ captures the strain-driven modification of magnetic preference relative to the unstrained structure, where positive (negative) values correspond to enhanced FM (A-AFM) stabilization.
}
\label{tab:NiNb4Se8_strain_rel}
\end{table*}